\documentclass[journal,10pt]{IEEEtran}
\usepackage{cite}
\usepackage{graphicx,color,epsfig,rotating}
\usepackage{amsfonts,amsmath,amssymb,bbm}
\usepackage{algorithm}
\usepackage{algpseudocode}
\usepackage{subfigure}
\usepackage{amsmath}
\usepackage{cite}
\usepackage{placeins}
\usepackage{graphicx}
\usepackage[latin1]{inputenc}
\usepackage{amssymb}
\usepackage{multirow}
\usepackage{stfloats}
\usepackage{tabularx} 
\usepackage{booktabs} 
\usepackage{url}
\usepackage{bm}
\usepackage{soul}
\usepackage{float}
\usepackage{lipsum}     
\usepackage{cuted} 
\usepackage{pstricks}
\usepackage{arydshln}
\usepackage{xspace} 

\allowdisplaybreaks

\long\def\comment#1{}

\makeatletter
\newcommand{\thickhline}{%
    \noalign {\ifnum 0=`}\fi \hrule height 1pt
    \futurelet \reserved@a \@xhline
}
\newcolumntype{"}{@{\hskip\tabcolsep\vrule width 1pt\hskip\tabcolsep}}
\makeatother

\makeatletter
\newcommand{\subalign}[1]{
  \vcenter{%
    \Let@ \restore@math@cr \default@tag
    \baselineskip\fontdimen10 \scriptfont\tw@
    \advance\baselineskip\fontdimen12 \scriptfont\tw@
    \lineskip\thr@@\fontdimen8 \scriptfont\thr@@
    \lineskiplimit\lineskip
    \ialign{\hfil$\m@th\scriptstyle##$&$\m@th\scriptstyle{}##$\crcr
      #1\crcr
    }%
  }
}
\makeatother

\newtheorem{example}{Example}
\newtheorem{theorem}{Theorem}
\newtheorem{lemma}{Lemma}
\newtheorem{corollary}{Corollary}
\newtheorem{remark}{Remark}

\def \inputsum{{\sum_{i=1}^KW_i}}
\def \rzsigma{{R_{Z_{\Sigma}}}}
\def \rzsigmastar{{R_{Z_{\Sigma}}^*}}
\def \rx{{R_X}} 

\def \rz{{R_Z}} 
\def \rs{{R_S}} 
\def \rsstar{{R_S^*}} 
\def \lx{{L_X}} 
\def \lz{{L_Z}} 
\def \ls{{L_S}}

\def \rzstar{{R_Z^*}}

\def  \lzsigma{{L_{Z_{\Sigma}}}}

\def  \KchooseG{{\binom{K}{G}}}
\def  \KchooseGminus{{\binom{K-1}{G-1}}}

\let\trm\textrm
\let\tbf\textbf
\let\tit\textit

\let\mbb\mathbb

\let \bksl\backslash
\let \ovst \overset

\newcommand{\wrt}{with respect to\xspace}

\newcommand{\corrd}{correlated\xspace}

\newcommand{\insum}{input sum\xspace}

\newcommand{\dsa}{decentralized secure aggregation\xspace}
\newcommand{\Dsa}{Decentralized secure aggregation\xspace}

\newcommand{\decen}{decentralized\xspace}

\newcommand{\reqd}{required\xspace}

\newcommand{\eg}{e.g.\xspace}

\newcommand{\ie}{i.e.\xspace}
\newcommand{\msg}{message\xspace}
\newcommand{\msgs}{messages\xspace}
\newcommand{\Wlog}{Without loss of generality\xspace}

\newcommand{\Ip}{In particular\xspace}
\newcommand{\af}{as follows\xspace}
\newcommand{\resp}{respectively\xspace}
\newcommand{\iid}{i.i.d.\xspace}

\newcommand{\Thm}{Theorem\xspace}

\newcommand{\info}{information\xspace}

\newcommand{\itic}{information-theoretic\xspace}

\newcommand{\etal}{\textit{et al.}\xspace}

\newcommand{\arbicorr}{arbitrarily correlated\xspace}

\newcommand{\grp}{group\xspace}

\newcommand{\grpws}{groupwise\xspace}

\newcommand{\agg}{aggregation\xspace}

\newcommand{\secagg}{secure aggregation\xspace}

\newcommand{\Fex}{For example\xspace}
\newcommand{\indep}{independent\xspace}

\newcommand{\indepce}{independence\xspace}

\newcommand{\indiv}{individual\xspace}
\newcommand{\Indiv}{Individual\xspace}
\newcommand{\comm}{communication\xspace}
\newcommand{\Comm}{Communication\xspace}

\newcommand{\achvblty}{achievability\xspace}

\newcommand{\distn}{distribution\xspace}

\newcommand{\muinfo}{mutual information\xspace}

\newfont{\bbb}{msbm10 scaled 700}

\newfont{\bb}{msbm10 scaled 1100}

\newcommand{\Am}{{\bf A}}
\newcommand{\Bm}{{\bf B}}

\newcommand{\Ac}{{\cal A}}
\newcommand{\Bc}{{\cal B}}

\newcommand{\Gc}{{\cal G}}

\newcommand{\Rc}{{\cal R}}
\newcommand{\Sc}{{\cal S}}
\newcommand{\Tc}{{\cal T}}

\newcommand{\eqdef}{\stackrel{\Delta}{=}}

\newcommand{\be}{\begin{equation}}
\newcommand{\ee}{\end{equation}}
\newcommand{\bea}{\begin{eqnarray}}
\newcommand{\eea}{\end{eqnarray}}

\newcommand{\bl}[1]{{\color{blue} #1}}
 
\newcommand{\xz}[1]{{\color{red} [XZ's COMMENT: #1]}}

\begin{document}

\title{The Capacity of Collusion-Resilient Decentralized Secure Aggregation with Groupwise Keys}

\author{
Zhou Li,~\IEEEmembership{Member,~IEEE},
Xiang~Zhang,~\IEEEmembership{Member,~IEEE},
Yizhou Zhao,~\IEEEmembership{Member,~IEEE},
Haiqiang Chen,~\IEEEmembership{Member,~IEEE},
Jihao Fan,~\IEEEmembership{Member,~IEEE},
and Giuseppe Caire,~\IEEEmembership{Fellow,~IEEE}

\thanks{Z. Li and H. Chen are with Guangxi Key Laboratory of Multimedia Communications and Network Technology, 
Guangxi University, Nanning 530004, China (e-mail: \{lizhou, haiqiang\}@gxu.edu.cn).}

\thanks{X. Zhang and G. Caire are with the Department of Electrical Engineering and Computer Science, Technical University of Berlin, 10623 Berlin, Germany (e-mail: \{xiang.zhang, caire\}@tu-berlin.de).
}

\thanks{Y. Zhao is with the College of Electronic and Information Engineering, Southwest University, Chongqing, China (e-mail: onezhou@swu.edu.cn).}

\thanks{J. Fan is with School of Cyber Science and Engineering, Nanjing University of Science and Technology, Nanjing 210094, China and also with Laboratory for Advanced Computing and Intelligence Engineering, Wuxi 214083, China (e-mail: jihao.fan@outlook.com).
}



}

\maketitle

\begin{abstract}
This paper investigates  the information-theoretic decentralized secure aggregation (DSA) problem  under practical groupwise secret keys and  collusion resilience.
In DSA, $K$ users are interconnected through error-free broadcast channels. Each user holds a private input and aims to compute the sum of all other users' inputs, while satisfying the security constraint that no user, even when colluding with up to $T$ other users, can infer any information about the inputs beyond the recovered sum. To ensure security, users are equipped with secret keys to mask their inputs.
Motivated by recent advances in efficient group-based key generation protocols, we consider the symmetric groupwise key setting, where every subset of $G$ users shares a group key that is independent of all other group keys. The problem is challenging because the recovery and security constraints must hold simultaneously for all users, and the structural constraints on the secret keys limit the flexibility of key correlations.
We characterize the optimal rate region consisting of all achievable pairs of per-user broadcast communication rate and groupwise key rate.
In particular,  we show that DSA with groupwise keys is infeasible when $G=1$ or $G\ge K-T$. Otherwise, when $2\le G<K-T$, to securely compute one symbol of the desired sum,  each user must broadcast at least one symbol, and each group key must contain at least 
$(K-T-2)/\binom{K-T-1}{G}$ independent symbols.
Our results establish the fundamental limits of DSA with groupwise keys and provide design insights for communication- and key-efficient secure aggregation in decentralized learning systems.
\end{abstract}


\section{Introduction}
\label{sec: intro} 
Federated learning (FL) has emerged as an important paradigm for privacy-preserving distributed machine learning, enabling multiple users to collaboratively train a global model without directly sharing their  local datasets~\cite{mcmahan2017communication,konecny2016federated,kairouz2021advances,yang2018applied}. In a typical centralized FL system, each user (or client) locally trains a model using its private dataset and then uploads the local update (e.g., model parameters or gradients) to a central server. The server aggregates the received updates -- typically in the form of weighted summation~\cite{mcmahan2017communication,karimireddy2020scaffold} -- to generate the global model, which is then redistributed to the users for the next training round. Recently, decentralized federated learning (DFL)~\cite{he2018cola,beltran2023decentralized} has also gained increasing attention due to its advantages in mitigating the single point of failure  of the \agg  server, as well as scalability, robustness, and trust among participants~\cite{wang2021edge,beltran2023decentralized}.

Although users in FL do not explicitly share their raw data, the exchange of local model updates still poses significant privacy risks~\cite{bouacida2021vulnerabilities, geiping2020inverting, mothukuri2021survey}. It has been shown that adversaries can exploit the shared gradients to infer sensitive information about users' private datasets through model inversion or membership inference attacks.
These threats have motivated the development of secure aggregation (SA) protocols~\cite{bonawitz2017practical, bonawitz2016practical, 9834981,wei2020federated,hu2020personalized,zhao2020local,yemini2023robust,so2021turbo, so2022lightsecagg, liu2022efficient,jahani2023swiftagg+}, which guarantee that the server learns only the aggregate of users' updates, without gaining access to any individual contribution.
Early work of Bonawitz~\etal~\cite{bonawitz2017practical} proposed cryptographic SA protocols that employ random seed-based pairwise masking among users to hide individual updates.
However, these methods provide only computational security guarantees.

On a different front, information-theoretic secure aggregation (SA)~\cite{9834981,zhao2023secure}  studies the fundamental performance  limits of communication and secret key generation  under  perfect security. Unlike cryptographic approaches, \itic SA provides unconditional security guarantees regardless of the adversary's computational power. In particular, Zhao and Sun~\cite{zhao2023secure} studied \secagg in a centralized  server-client architecture where the server  aims to  compute  the sum of the \emph{inputs} -- abstraction of the local model updates in FL -- while being prevented  from learning anything about  the inputs beyond their sum. It was shown that, to securely compute one symbol of the desired sum, each user must hold at least one secret key symbol, transmit at least one symbol to the server, and all users must collectively hold at least $K-1$ ($K$  is the number of users) independent key symbols. This result establishes the fundamental limits of SA in centralized systems. 

Recent progress in information-theoretic SA has extended its applicability by addressing a variety of practical challenges, including user dropout and  collusion resilience~\cite{9834981,jahani2022swiftagg,jahani2023swiftagg+},  selective user participation~\cite{zhao2022mds, zhao2023optimal,zhao2024secure}, heterogeneous security constraints~\cite{li2023weakly,li2025weakly,li2025collusionresilienthierarchicalsecureaggregation}, oblivious-server settings~\cite{sun2023secure}, multiple-objective recovery~\cite{yuan2025vector}, hierarchical SA (HSA)~\cite{zhang2024optimal,10806947,egger2024privateaggregationhierarchicalwireless,zhang2025fundamental,11195652,li2025collusionresilienthierarchicalsecureaggregation,egger2023private, lu2024capacity}, and decentralized SA (DSA)~\cite{Zhang_Li_Wan_DSA}.
For example, weak security models that protect only subsets of users were studied in~\cite{li2023weakly,li2025weakly,li2025collusionresilienthierarchicalsecureaggregation}. SA with user selection was investigated in~\cite{zhao2022mds,zhao2023optimal,zhao2024secure} to capture the effect of partial participation in federated learning systems. Moreover,  motivated by \decen FL~\cite{he2018cola,beltran2023decentralized},   DSA~\cite{Zhang_Li_Wan_DSA} studies SA in a fully \decen network topology where all users aim to recover the  global sum \emph{simultaneously} through inter-user broadcast transmission. Compared with centralized SA, where the server is untrusted and must not learn any individual input, DSA assumes that \emph{every} user is untrusted; in particular, the inputs of any set of $K-1 $ users must remain hidden from the remaining user.

An important variant of SA is the use of \textit{groupwise} keys~\cite{zhao2023secure,wan2024information,wan2024capacity}, as they allow each subset of $G$ users to share a
common group key that is \indep of other group keys. \Fex, $G=2$ corresponds to the popular pairwise key setting~\cite{bonawitz2017practical, so2022lightsecagg}. 
In practice, such keys  can be  generated efficiently via inter-user \comm with \itic \cite{maurer2002secret, ahlswede2002common} or computational~\cite{Wallez2025TreeKEMAM,Rescorla2008TheTL} security guarantees, thereby removing  the need  for a dedicated key \distn server  that is otherwise required when \emph{arbitrarily} correlated keys are used~\cite{9834981, zhang2025fundamental, Zhang_Li_Wan_DSA}.
For centralized SA, it was shown that~\cite{zhao2023secure}  \secagg is infeasible if  $G = 1$ or $G > K - T$ ($T$ denotes the maximum number of allowed colluding users). 
Otherwise,  to securely compute one symbol of the desired sum, each user must transmit at least  one symbol to the server, and each group key must contain at least $(K - T - 1)/\binom{K - T}{G}$ \indep symbols. This  result reveals a fundamental trade-off  between the \grp key  size  and the
group granularity $G$  as well as   the collusion tolerance $T$. However, in decentralized settings that employ groupwise keys, the optimal communication and key rates remain unknown.

Motivated by the increasing    security demands in decentralized federated learning, and by the practical advantages of groupwise secret keys, we study \emph{decentralized secure aggregation (DSA) with groupwise keys and collusion resilience}. Specifically, we consider secure aggregation over a fully-connected network of $K$ users, where each user holds a private input $W_k$ and is connected to the remaining $K-1$ users through an error-free broadcast channel, as illustrated in Fig.~\ref{fig:model}. 
Our objective is to allow every user to recover the global input sum $\sum_{k=1}^KW_k$, while satisfying the security requirement that any user $k$ must not learn any information about the other users' inputs $\{W_i\}_{i \ne k}$ beyond their aggregate sum and what can be inferred through collusion with up to $T$ other users.
To protect the inputs, the \grpws keys are  adopted where each subset of  $G$ users share a common \grp key $S_\Gc$ that is \indep of other \grp keys, and all user collectively hold $\{S_\Gc\}_{\Gc \in \binom{[K]}{G}}$. Based on the input $W_k$ and the stored keys  $\left\{S_{\mathcal{G}}\right\}_{ \Gc \in \binom{[K]}{G},  k \in \mathcal{G}}$, user $k$ generates a message $X_k$ that is broadcast to all other users. Then, upon receiving  the \msgs $\{X_i\}_{i\in [K]\bksl \{k\} }$, each user $k$ should correctly recover $\sum_{i\in [K]\bksl \{k\}  }W_i$ while satisfying the security constraints.

Our main result establishes the complete capacity region of decentralized secure aggregation with groupwise keys. Specifically, we show that DSA is infeasible when $G = 1$ or $G \ge K - T$. In all other cases, each user must transmit at least one symbol, and each group of users must hold at least 
$(K - T - 2)/\binom{K - T - 1}{G}$ 
independent key symbols to securely compute the global sum. These results characterize the fundamental limits of DSA in fully connected networks and show that the design of groupwise keys directly determines both the feasibility and efficiency of collusion-resilient aggregation.

\subsection{Summary of Contributions}
The main contributions of this paper are summarized as follows:
\begin{itemize}
    \item We present an information-theoretic model of decentralized secure aggregation (DSA) with groupwise keys under user collusion. The model explicitly captures how the group size $G$ constrains the maximum number of colluding users that the system can tolerate, extending the conventional centralized secure summation framework to fully decentralized settings.

    \item We propose a structured \emph{key-neutralization} scheme that achieves perfect security with minimum communication and groupwise key rates. The scheme carefully coordinates groupwise keys to neutralize potential information leakage from colluding users.

    \item We prove a matching converse using tight information-theoretic bounds, establishing the exact capacity region of DSA. This result identifies the fundamental limits on communication and key efficiency in decentralized secure aggregation.
\end{itemize}

\textit{Paper Organization.} 
The rest of the paper is organized as follows. Section~\ref{sec: problem description} formulates the DSA problem with groupwise keys and collusion. Section~\ref{sec:main result} presents the main results, which includes the characterization of the optimal rate region. Section~\ref{sec: ach scheme} illustrates two concrete examples to convey the achievability idea, followed by the general code construction and proof.  The general converse proof is presented  in Section~\ref{sec: converse}  Finally, Section~\ref{sec:conclusion & future directions} concludes the paper.

\tit{Notation.}  For integers $m\le n$, $[m:n] \eqdef \{m,m+1, \cdots, n\}$ if $m\le n$, and $[m:n]=\emptyset$ if $m>n$. $[n]\eqdef\{1,\cdots,n\}$.
For  $X_1,\cdots, X_n$, denote $X_{1:n}\eqdef\{X_1,\cdots, X_n\}$. 
Bold capital letters (\eg, $\Am, \Bm)$ and calligraphic letters (\eg, $\Ac, \Bc$) represent    matrices and sets, \resp.
$\Ac \bksl \Bc\eqdef \{x\in \Ac: x\notin \Bc\}$. Moreover, $\binom{\Ac}{n}\eqdef \{ \Sc \subseteq \Ac : |\Sc|=n  \}$ denotes all the $n$-element subsets of $\Ac$.

\section{Problem Statement}
\label{sec: problem description}
\begin{figure}[ht]
    \centering
    \includegraphics[width=0.45\textwidth]{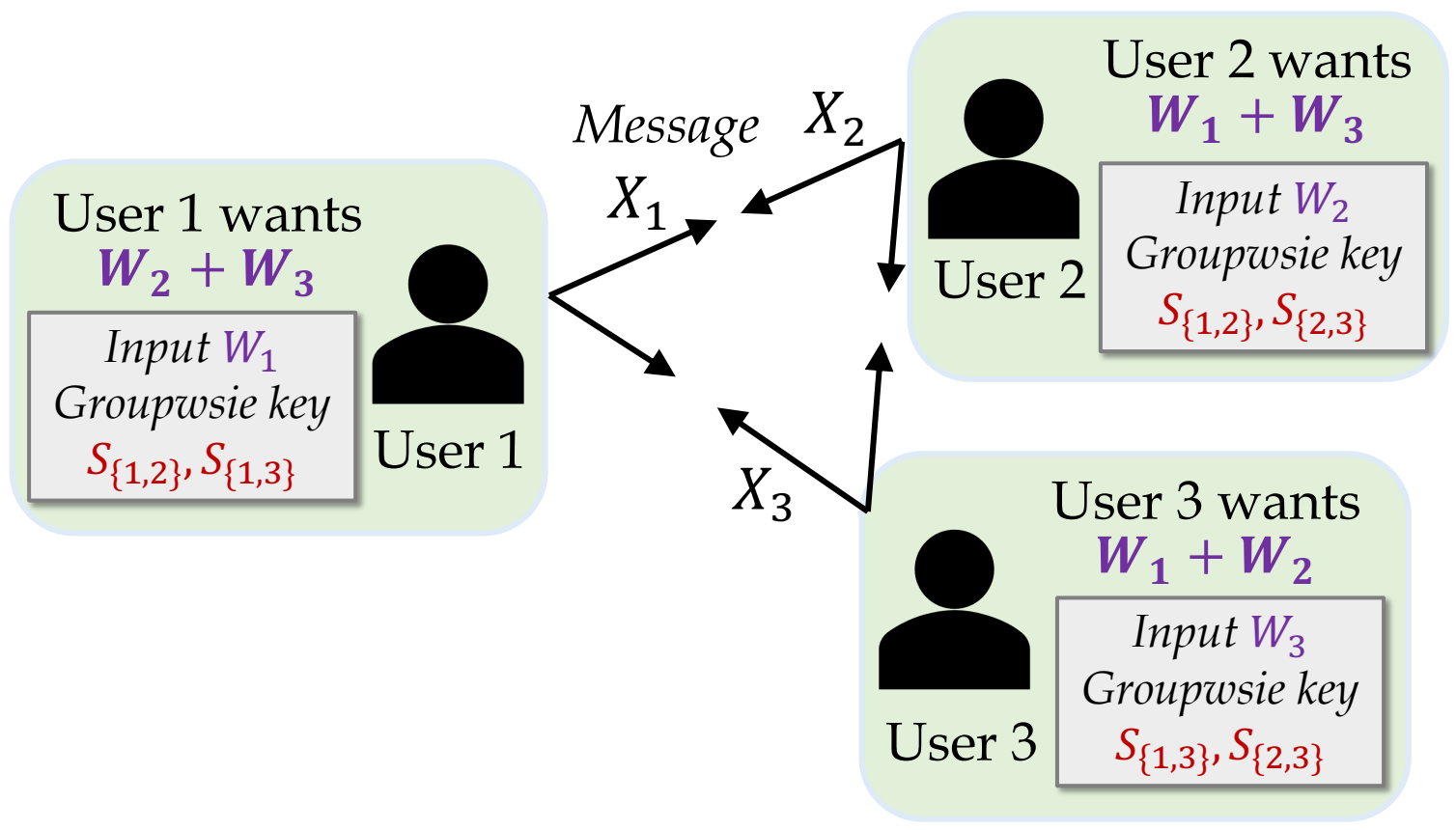}
    \caption{\Dsa with 3 users. User 1 aims to recover the sum $W_2 + W_3$ from the received messages $X_2$ and $X_3$ and its own information $W_1$ and groupwise key $S_{\{1,2\}},S_{\{1,3\}}$, while being prevented from learning any additional information about the pair of inputs $(W_2, W_3)$ beyond their sum. The same security requirement is imposed on the other two users as well.  }
    \label{fig:model}
\end{figure}

The decentralized secure aggregation involves $K\geq 3$ users, where each user is connected to each other through an error-free broadcast channel as shown in Fig.~\ref{fig:model}.
Each user \(k \in [K]\) holds an \emph{input variable} \(W_k\) consisting of \(L\) i.i.d.\ symbols uniformly distributed over a finite field \(\mathbb{F}_q\). The inputs $W_{1:K}$ are assumed to be independent of each other\footnote{\label{footnote:input iid uniformity}It should be emphasized that the assumptions of input uniformity and independence are only needed for the converse proof establishing optimality. The proposed scheme itself guarantees security for arbitrary input distributions and correlations. See Sections \ref{sec: ach scheme} and \ref{sec: converse}  for details.}, \ie,  
\begin{align}
\label{eq:input independence}
H(W_{1:K}) &=\sum_{k=1}^KH(W_k),\notag\\
H(W_k) & = L\, \trm{(in $q$-ary unit)},\; \forall  k\in[K].
\end{align} 
To ensure security, users must share certain secrets (i.e., keys) that are independent of their input vectors. In this work, we adopt a \textit{groupwise key} structure, where each key is shared among a group of $G \in [K] $ users, and all keys are mutually independent. Specifically, let $\binom{K}{G}$ denote the number of possible user groups of size $G$. For each group $\mathcal{G} \in \binom{[K]}{G}$, we assign an independent random variable $S_{\mathcal{G}}$, which consists of $L_S$ i.i.d. symbols drawn uniformly from $\mathbb{F}_q$. Therefore, 
\begin{align}
H\left(\left(S_{\mathcal{G}}\right)_{\mathcal{G} \in \binom{[K]}{G}},W_{1:K}\right)
&= \sum_{\mathcal{G} \in \binom{[K]}{G}} H(S_{\mathcal{G}})+\sum_{k\in [K]}H\left(W_k\right)\notag\\
&= \binom{K}{G} L_S+\sum_{k\in [K]}H\left(W_k\right). \label{ind_S}
\end{align}
$S_{\mathcal{G}}$ is then shared exclusively by users in $\Gc$. The \indiv key variable $Z_k$ of user  $k$ is  the set of all \grp keys assigned to it, \ie, 
\begin{eqnarray}
\label{eq:def Zk}
	Z_k=\left\{S_{\mathcal{G}}\right\}_{\forall \Gc \in \binom{[K]}{G} \trm{ s.t. }   k \in \mathcal{G}}, \; k \in [K]. \label{Z_individual}
\end{eqnarray}

To aggregate the inputs, each user $k\in[K]$ generates a message  $X_k$  of length $L_X$ symbols from $\mathbb{F}_q$, using its own input $W_k$ and key $Z_k$. The message $X_k$ is then broadcast to the remaining $K-1$ users. Specifically, we assume $X_k$ is a deterministic function of $W_k$ and $Z_k$,  so that
\begin{eqnarray}
\label{eq:H(Xk|Wk,Zk)=0}
    H(X_k|W_k,Z_k)=0, \forall  k\in[K]. \label{message}
\end{eqnarray}
Each user $k\in[K]$, by combining the messages received from others with its own input and secret key, should be able to recover the sum of all inputs.
\begin{align}
\label{eq:recovery constraint}
& \mathrm{[Recovery]} \quad H\left(\inputsum \bigg|\{X_i\}_{i\in [K]\bksl \{ k\} }, W_k,Z_k\right  )=0,\notag\\
& \hspace{6.5cm}\forall  k\in[K].
\end{align}
It is worth noting that $X_k$ does not appear in the conditioning terms, as it is uniquely determined by $W_k$ and $Z_k$. In addition, since $W_k$ is locally available to user $k$, it suffices for the user to reconstruct $\sum_{i \in [K]\setminus \{k\}} X_i$, from which the overall sum can be subsequently derived.

\tbf{Security model.}
We impose the security constraint that each user's input should remain hidden to other users (except for the case of  collusion) during the broadcast  \msg exchange.  \Ip, it is \reqd that  each user  $k$, even colluding with at most  most $T$ other users $\Tc \subset [K]\bksl \{k\}$ where  $|\Tc|\le T$ -- \ie, user $k$ gains access to  the inputs and keys $\{W_i, Z_i\}_{i\in \Tc}$, should be  prevented from inferring anything about the inputs of the remaining users  $\{W_i\}_{ i \in [K]\bksl (\Tc \cup \{k\})}$. This security constraint  can be written in terms of mutual \info  as
\begin{align}
\label{eq:security constraint}
& \mathrm{[Security]}  I\Bigg( \{X_i\}_{i \in [K]\bksl \{k\}}; \{W_i\}_{i \in [K]\bksl \{k\}} \Big|\sum_{i=1}^K W_i, 
 W_k, Z_k,\notag\\
&  \{W_i, Z_i\}_{i\in \mathcal{T}}   \Bigg)  =0, \forall \Tc \subset [K]\bksl \{k\}, |\Tc|\le T, \forall k\in[K].
\end{align}

\tbf{Performance metrics.}
We study both the communication and secret key generation efficiency of DSA under the groupwise key setting. In particular, the communication rate, denoted by $\rx$, characterizes the number of transmitted symbols per message normalized by the input size, and is defined as
\begin{eqnarray}
    R_X \eqdef \lx/L.
\end{eqnarray} 
The groupwise key rate $\rs$ characterizes the number of symbols contained in each groupwise key (normalized  by the input size), which is defined as   
\begin{eqnarray}
    R_S \eqdef \ls/L,
\end{eqnarray}
where $L_S$ represents the number of symbols contained in $S_{\mathcal{G}}$. 
A rate tuple $(R_X,R_S)$ is said to be achievable if there is a decentralized secure aggregation scheme, for which the correctness constraint (\ref{eq:recovery constraint}) and the security constraint (\ref{eq:security constraint}) are satisfied, and the communication and groupwise key rates are no greater than $R_X$ and $R_S$, respectively. The closure of the set of all achievable rate tuples is called the optimal rate region (i.e., capacity region), denoted as $\mathcal{R}^*$.

\begin{remark}[\Indiv and Source Key Rates]
\label{remark:indiv and source key rates}
Since  each user's \indiv key $Z_k$ contains  $\KchooseG$ \grp keys which are mutually \indep by assumption (see (\ref{eq:def Zk})), the  total  number of  \indep symbols  in each $Z_k$  and in the collection  $Z_{1:K}$ is equal to $\lz\eqdef \KchooseGminus \ls$ and $\lzsigma \eqdef \KchooseG \ls $, \resp. 
Hence, the \indiv key rate $\rz \eqdef \lz/L$ 
and the  source key rate $\rzsigma \eqdef \lzsigma/L$ has a  deterministic relation with $\rs$, namely, 
\be 
\label{eq:relation Rz,Rzsigma,Rs}
\rz= \KchooseGminus\rs, \; 
\rzsigma =  \KchooseG \rs.
\ee
Therefore, the rate tuple $(\rx, \rz, \rzsigma)$ commonly used in the SA literature with arbitrarily \corrd keys~\cite{9834981, zhang2025fundamental,Zhang_Li_Wan_DSA} can be equivalently represented by $(\rx, \rs)$.
\end{remark}

\section{Main Result}
\label{sec:main result}
This section presents the main result of the paper, which  characterizes the optimal groupwise key rate for the proposed \dsa problem.

\begin{theorem}
\label{thm:main result}
\tit{For \dsa   with $K\ge  3$ users, at most $T \in [0:K-3]$ colluding users\footnote{The DSA problem is inherently infeasible when  there are less than 3 users, or more than $K-3$ colluding users. See a detailed explanation in~\cite{Zhang_Li_Wan_DSA}.},  and groupwise key group size $G\in [1:K]$, the optimal rate region is $\Rc^*=\emptyset$ (\ie, infeasible) if $G=1$ or $G\ge K-T$. Otherwise, when $2\le  G < K-T $, the optimal rate region is given by}
\be 
\label{eq:opt region,thm}
\Rc^*=\left\{
\left(R_X, R_{S}\right): \rx \ge 1, R_{S} \geq \cfrac{K-T-2}{\binom{K-T-1}{G}}
\right\}.
\ee 
\end{theorem}

The achievability and converse proof of Theorem~\ref{thm:main result} is presented in Sections~\ref{sec: ach scheme} and \ref{sec: converse}, \resp.
We highlight the implications  of the \Thm~\ref{thm:main result} \af:

\subsubsection{Infeasibility} 
When $G = 1$ or $G \ge K - T$, the DSA problem becomes infeasible, which corresponds to the infeasibility of input sum recovery and security, \resp. \Ip, when $G=1$, each user forms a group by itself and possesses an independent key. In this case, there is no correlation among $Z_{1:K}$, which is a necessary condition for key cancellation and, hence, for recovering the input sum.
When $G \ge K-T$, there are two cases after removing the colluded keys from any $|\Tc|=T$ users: 1) If $G=K-T$, the only key left is $S_{[K]\bksl \Tc  }$ which is shared among the non-colluding users; 2) If $G>K-T$, there  is no key left in the  non-colluding  users. In both cases, security is impossible.

\subsubsection{Interpretation of $\rs$} When feasible, the minimum \grp key rate is equal to
$R_S^*=\frac{(K-T-2)}{\binom{K-T-1}{G}}, \forall G \in[2:K-T-1]  $. An intuitive explanation of  $R_S^*$ is \af. Consider  the \agg  process at any user $k$ and the worst-case that it colludes with $|\Tc|=T$  other users. After eliminating the  group keys $\{S_\Gc  :\forall \Gc, \Gc \cap (\Tc \cup\{k\}) \ne \emptyset    \}   $ known to user $k$, and the inputs $\{W_i\}_{i\in  \Tc\cup \{k\} }$, the recovery at user $k$ becomes equivalent to a centralized SA system  where user $k$ serves as the server who aims to compute the sum of the inputs from users $ [K]\bksl (\Tc \cup \{k\}  ) $. By the fundamental result of Zhao and Sun \cite{zhao2023secure},  to achieve security, the $K-T-1$ users in  $ [K]\bksl (\Tc \cup \{k\}  ) $ must collectively hold at least $(K-T-2)L$ \indep  key symbols. Since there are $\binom{K-T-1}{G}$ \grp keys owned exclusively by these users, the size of each \grp key needs to be at least 
$\ls\ge\frac{(K-T-2)L}{\binom{K-T-1}{G}}$, implying 
$\rs \ge \frac{K-T-2}{\binom{K-T-1}{G} } $. We formalize these intuitions as rigorous entropic converse proof and a matching \achvblty scheme in Sections \ref{sec: converse} and \ref{sec: ach scheme}, \resp.

\begin{figure}[t]
    \centering
    \includegraphics[width=0.45\textwidth]{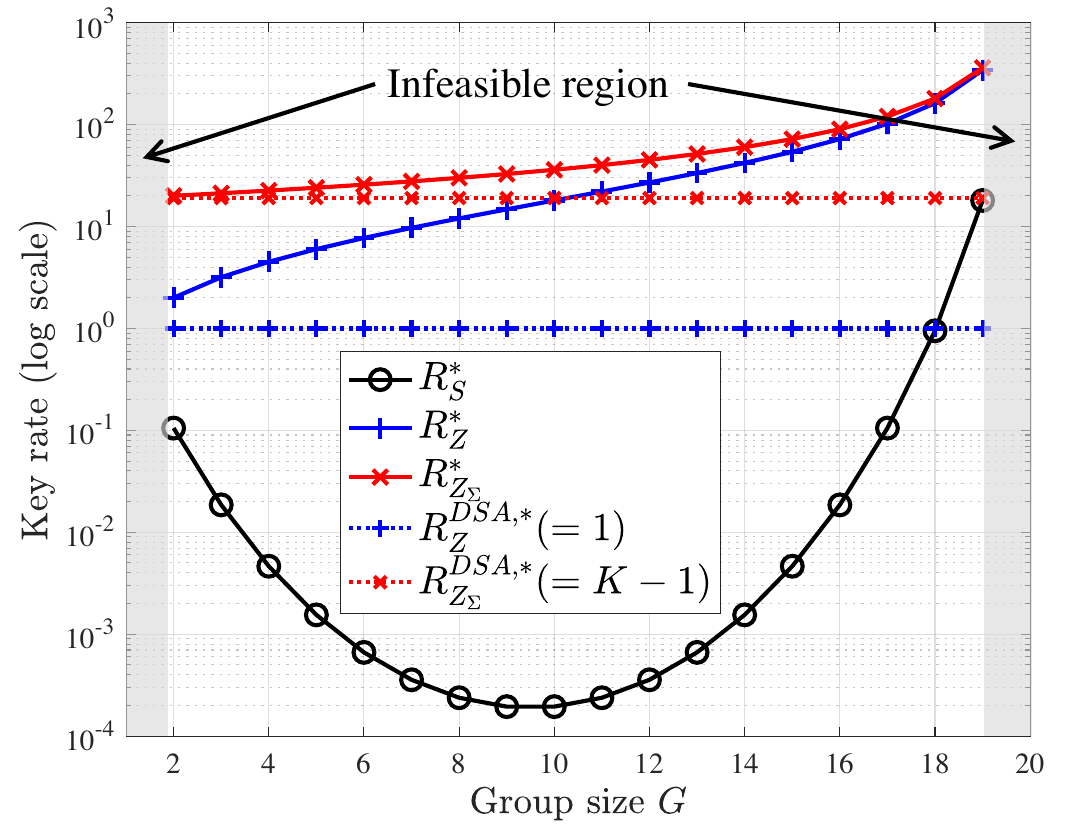}
    \caption{Comparison of \indiv and source key rates under the \grpws  key setting (this paper) with \arbicorr keys (Zhang \etal~\cite{Zhang_Li_Wan_DSA}) for $K=20$ and $T=0$.  }
    \label{fig:fig_rate}
    \vspace{-.5cm}
\end{figure}

\subsubsection{Impact of Group Size $G$} 
The optimal key rate  $\rsstar$ varies as the group size $G$ changes. Fig. \ref{fig:fig_rate} shows the optimal \grp key, \indiv key and source key rates $\rsstar, \rzstar $ and $\rzsigmastar$ ($\rzstar$ and 
$\rzsigmastar$ are computed based on $\rsstar$ according to (\ref{eq:relation Rz,Rzsigma,Rs})), as functions of  $G$ for $K=20, T=0$. For comparison, the \indiv and source key rates $R_Z^{DSA,*}=1$  and $R_{Z_\Sigma}^{DSA, *}=K-1$ of  
the DSA scheme in~\cite{Zhang_Li_Wan_DSA} with \emph{arbitrarily correlated} keys are also plotted.
Several observations can be made immediately. First, $\rzstar$  and $\rzsigmastar$ are monotonically increasing \wrt $G$. By (\ref{eq:relation Rz,Rzsigma,Rs}) and (\ref{eq:opt region,thm}), we  have
$
\rzstar= \frac{(K-1)\cdots(K-T)(K-T-2) G}{(K-G)\cdots (K-G-T)}$ and $\rzsigmastar=\frac{K}{G}\rzstar$, both of which are increasing functions of $G$.  
This monotonicity can be explained as follows. As $G$ increases, the amount of shared randomness among users' keys also increases. To maintain a sufficient number of independent key symbols at each user -- necessary for protecting the inputs -- the size of the individual key must  increase accordingly.
Second, as $G$ increases, $\rsstar$ first decreases and then increases, attaining its minimum at $G^*= \left\lfloor \frac{K-T-1}{2} \right\rfloor$. 
Intuitively, this behavior arises  because the number of \grp keys available for \secagg at any user $k$ -- after excluding the keys  contained in $Z_k$ and $\{Z_i\}_{i\in  \Tc}$ -- is given by
\be  
\left| \left\{  S_\Gc\right\}_{\Gc:\Gc  \cap (\Tc \cup \{k\}  )=\emptyset, |\Tc|=T }   \right| = \binom{K-T-1}{G}.
\ee
This quantity is maximized when $G^*= \left\lfloor \frac{K-T-1}{2} \right\rfloor$, which offers maximal flexibility in key selection and  thereby minimizes  the  required \grp key size.
Lastly, DSA with groupwise keys requires higher individual and source key rates than the case with arbitrarily correlated keys, highlighting the additional key rate cost incurred by enforcing structural constraints on the secret keys.

\section{Achievable Scheme}
\label{sec: ach scheme}
 Before describing the general scheme, 
 we first  present two motivating examples to highlight the ideas behind the proposed design. As already  mentioned in Footnote \ref{footnote:input iid uniformity}, the proposed \secagg scheme ensures security and correctness for any arbitrary input \distn. The \iid uniformity  assumption in (\ref{eq:input independence}) is only necessary for establishing the optimality via  the converse proof in Section \ref{sec: converse}.

\subsection{Motivating Examples}
\label{subsec:examples}

\begin{example}[$K=3, T=0, G=2$] 
\label{example: ach scheme1}
Consider the case with $K=3$ users where user $k\in [3]$ does not collude with any other users (i.e., $T=0$), as shown in Fig.~\ref{fig:model} (Section \ref{sec: problem description}). Every $G=2$ users share a groupwise key. The rate to be achieved is  $R_X=1, R_S=(K-T-2)/\binom{K-T-1}{G}=1/1$. Set $L=L_X=1$, i.e., $W_1, W_2, W_3$; Set $L_S=1$, i.e., $S_{\{1,2\}}=A, S_{\{1,3\}}=B, S_{\{2,3\}}=C$. Then $Z_1=\{A,B\}, Z_2=\{A,C\}, Z_3=\{B, C\}$. 
The message are
\begin{eqnarray}
    X_1&=&W_1+A+B,\notag\\
    X_2&=&W_2-A+C,\notag\\
    X_3&=&W_3-B-C, \label{ex1:scheme}
\end{eqnarray}
where each $W_k$ contains 1 symbol from $\mbb{F}_q$ and $A,B,C$ are three i.i.d. uniform key symbols from $\mbb{F}_q$, where $q=2$. With the code assignment completed, we next analyze the correctness and security of the proposed scheme. For correctness, each user must be able to recover the global sum $W_1+W_2+W_3$. Consider user 1 as an example. With its own input $W_1$ and key $Z_1$, and given the received messages $X_2$ and $X_3$, user 1 should be able to reconstruct $W_2+W_3$, and hence obtain the complete sum:
\begin{align}
&X_2+X_3+\underbrace{A+B}_{Z_1}\notag\\
  \overset{(\ref{ex1:scheme})}{=}&  (W_2-A+C)+ (W_3-B-C) +A+B\notag\\
=&W_2+W_3.
\end{align}
Similarly, the two remaining users can also recover the desired \insum \af:
\begin{align}
\trm{User  2:}\quad X_1+X_3-\underbrace{A+C}_{Z_2} & = W_1 + W_3,  \notag\\
\trm{User  3:}\quad X_1+X_2-\underbrace{B-C}_{Z_3} & = W_1 + W_2.
\end{align}

We now show that  the proposed scheme satisfies the security constraint (\ref{eq:security constraint}). \Wlog, let us consider User 1. By (\ref{eq:security constraint}), we have
\begin{align}
& I\left(X_2,X_3;W_2,W_3|W_2+W_3,W_1, Z_1\right)\notag\\
=&  H(X_2,X_3|W_2+W_3,W_1, Z_1) \notag\\
&- H(X_2,X_3|W_2,W_3,W_1, Z_1).\label{eq:exmpl1}\\
\overset{(\ref{ex1:scheme})}{=} &H\left(W_2-A+C,W_3-B-C |W_2+W_3,W_1, A, B\right)  \notag\\
&-H\left(W_2-A+C,W_3-B-C |W_2,W_3,W_1, A, B\right)  \notag\\
=& H(W_2+C,W_3-C |W_2+W_3, W_1,A, B)\notag\\
&-H(C |W_2,W_3, W_1,A, B)\notag\\
=&H(W_2+C,W_3-C |W_2+W_3)-H(C)\label{eq:ex1pr1}\\
=&H(W_2+C,W_3-C,W_2+W_3)-H(W_2+W_3)-1\\
=&H(W_2+C,W_3-C)-H(W_2+W_3)-1\label{eq:ex1pr2}\\
\leq&H(W_2+C)+H(W_3-C)-H(W_2+W_3)-1\label{eq:ex1pr3}\\
=&1+1-1-1=0.
\end{align}
Here, (\ref{eq:ex1pr1}) holds since $W_1, W_2, W_3, A, B, C$ are i.i.d., and (\ref{eq:ex1pr2}) holds because $W_2+W_3$ can be derived from $W_2+C$ and $W_3-C$. This completes the security proof for User 1; the cases for the other users follow analogously.
\hfill $\lozenge$
\end{example}

While the above example illustrates the basic idea of decentralized secure aggregation, it does not account for the possibility of user collusion. In the following, we extend the analysis to explicitly consider scenarios where multiple users may collude, and examine how such collusion affects the design and performance of decentralized secure aggregation.
\begin{example}[$K=5, T=1, G=2$]
\label{example: ach scheme2}
Consider a system with $K=5$ users, where at most $T=1$ user may collude with any given user $k \in [5]$. For a group size of $G=2$, there are ten groupwise keys: $S_{\{1,2\}}, S_{\{1,3\}},$ $ S_{\{1,4\}}, S_{\{1,5\}},
S_{\{2,3\}}, S_{\{2,4\}}, S_{\{2,5\}},
S_{\{3,4\}}, S_{\{3,5\}}, S_{\{4,5\}}$. The target communication and key rates are given by
$
R = 1,  R_S = (K - T - 2)/{\binom{K - T - 1}{G}} = \frac{2}{3}.$
Let the message length be $L = L_X = 3$, such that each user holds
\[
W_k = [W_k(1),\, W_k(2),\, W_k(3)]^{\mathsf{T}} \in \mathbb{F}_q^{3\times 1}, \quad k \in [5].
\]
Each groupwise key has length $L_S = 2$, i.e.,
\[
S_{\mathcal{G}} = [S_{\mathcal{G}}(1),\, S_{\mathcal{G}}(2)]^{\mathsf{T}} \in \mathbb{F}_q^{2\times 1}, \quad \mathcal{G} \subset [5],\, |\mathcal{G}| = 2.
\]

The messages are set as 
\begin{align}
     X_1 =&W_1 +{\bf H}_{\{1,2\}}^{3\times 2}S_{\{1,2\}} +{\bf H}_{\{1,3\}}S_{\{1,3\}}\notag\\
     &+{\bf H}_{\{1,4\}}S_{\{1,4\}}+{\bf H}_{\{1,5\}}S_{\{1,5\}}, \notag\\
     X_2 =& W_2-{\bf H}_{\{1,2\}}S_{\{1,2\}}+{\bf H}_{\{2,3\}}S_{\{2,3\}}\notag\\
     &+{\bf H}_{\{2,4\}}S_{\{2,4\}}+{\bf H}_{\{2,5\}}S_{\{2,5\}},\notag\\
     X_3 =& W_3-{\bf H}_{\{1,3\}}S_{\{1,3\}}-{\bf H}_{\{2,3\}}S_{\{2,3\}}\notag\\
     &+{\bf H}_{\{3,4\}}S_{\{3,4\}}+{\bf H}_{\{3,5\}}S_{\{3,5\}}, \notag\\
     X_4 =& W_4-{\bf H}_{\{1,4\}}S_{\{1,4\}}-{\bf H}_{\{2,4\}}S_{\{2,4\}}\notag\\
     &-{\bf H}_{\{3,4\}}S_{\{3,4\}}+{\bf H}_{\{4,5\}}S_{\{4,5\}}, \notag\\
     X_5 =& W_5-{\bf H}_{\{1,5\}}S_{\{1,5\}}-{\bf H}_{\{2,5\}}S_{\{2,5\}}\notag\\
     &-{\bf H}_{\{3,5\}}S_{\{3,5\}}-{\bf H}_{\{4,5\}}S_{\{4,5\}},\label{eq:hd2}
\end{align}
where each ${\bf H}_{\mathcal{G}}\in \mathbb{F}_q^{3 \times 2}$ is a $3 \times 2$ key precoding matrix.
Security can be guaranteed provided that the matrices ${\bf H}_{\mathcal{G}}$ are sufficiently generic. In particular, we will show in the general proof that when the matrices ${\bf H}_{\mathcal{G}}$ are randomly selected from a sufficiently large field, a secure construction exists. Although the field size $q$ of $\mathbb{F}_q$ is fixed, we can effectively enlarge the operating field by increasing the input block length $L$, thereby working over an appropriate extension field. For instance, when $q = 5$, we may choose
\begin{align}
     {\bf H}_{\{1,2\}}= \left[
     \begin{array}{cc}
          2&3\\
          4&0\\
          2&1
     \end{array}\right],
     {\bf H}_{\{1,3\}}= \left[
     \begin{array}{cc}
          3&2\\
          3&2\\
          3&0
     \end{array}\right],~\notag\\
     {\bf H}_{\{1,4\}}= \left[
     \begin{array}{cc}
          3&3\\
          1&0\\
          3&2
     \end{array}\right],
     {\bf H}_{\{1,5\}}= \left[
     \begin{array}{cc}
          3&4\\
          4&4\\
          3&0
     \end{array}\right],~\notag\\
     {\bf H}_{\{2,3\}}= \left[
     \begin{array}{cc}
          2&1\\
          0&1\\
          3&1
     \end{array}\right],
     {\bf H}_{\{2,4\}}= \left[
     \begin{array}{cc}
          0&0\\
          0&4\\
          4&2
     \end{array}\right],~\notag\\
     {\bf H}_{\{2,5\}}= \left[
     \begin{array}{cc}
          1&0\\
          0&1\\
          4&0
     \end{array}\right],
     {\bf H}_{\{3,4\}}= \left[
     \begin{array}{cc}
          4&2\\
          3&3\\
          2&0
     \end{array}\right],~\notag\\
     {\bf H}_{\{3,5\}}= \left[
     \begin{array}{cc}
          0&0\\
          0&1\\
          1&3
     \end{array}\right],
     {\bf H}_{\{4,5\}}= \left[
     \begin{array}{cc}
          0&3\\
          1&1\\
          2&0
     \end{array}\right].\label{V_example2}
\end{align}

With the code assignment complete, we now analyze the correctness and security of the proposed scheme. 
For correctness, each user must be able to recover the global sum \( W_1 + W_2 + W_3 + W_4 + W_5 \). 
Take user~1 as an example. Given its own input \( W_1 \) and key \( Z_1 \), as well as the received messages 
\( X_2, X_3, X_4, \) and \( X_5 \), user~1 should be able to reconstruct \( W_2 + W_3 + W_4 + W_5 \), 
and thus obtain the complete sum.
\begin{align}
&X_2+X_3+X_4+X_5+\underbrace{{\bf H}_{\{1,2\}}^{3\times 2}S_{\{1,2\}} +{\bf H}_{\{1,3\}}S_{\{1,3\}}}_{Z_1}\notag\\
&+\underbrace{{\bf H}_{\{1,4\}}S_{\{1,4\}}+{\bf H}_{\{1,5\}}S_{\{1,5\}}}_{Z_1}\notag\\
  \overset{(\ref{eq:hd2})}{=}&  W_2-{\bf H}_{\{1,2\}}S_{\{1,2\}}+{\bf H}_{\{2,3\}}S_{\{2,3\}}+{\bf H}_{\{2,4\}}S_{\{2,4\}}\notag\\
  &+{\bf H}_{\{2,5\}}S_{\{2,5\}}+W_3-{\bf H}_{\{1,3\}}S_{\{1,3\}}-{\bf H}_{\{2,3\}}S_{\{2,3\}}\notag\\
  &+{\bf H}_{\{3,4\}}S_{\{3,4\}}+{\bf H}_{\{3,5\}}S_{\{3,5\}} +W_4-{\bf H}_{\{1,4\}}S_{\{1,4\}}\notag\\
  &-{\bf H}_{\{2,4\}}S_{\{2,4\}}-{\bf H}_{\{3,4\}}S_{\{3,4\}}+{\bf H}_{\{3,5\}}S_{\{3,5\}}\notag\\
  &+W_5-{\bf H}_{\{1,5\}}S_{\{1,5\}}-{\bf H}_{\{2,5\}}S_{\{2,5\}}-{\bf H}_{\{3,5\}}S_{\{3,5\}}\notag\\
     &-{\bf H}_{\{4,5\}}S_{\{4,5\}}+{\bf H}_{\{1,2\}}^{3\times 2}S_{\{1,2\}} +{\bf H}_{\{1,3\}}S_{\{1,3\}}\notag\\
     &+{\bf H}_{\{1,4\}}S_{\{1,4\}}+{\bf H}_{\{1,5\}}S_{\{1,5\}}\\
=&W_2+W_3+W_4+W_5.
\end{align}
Similarly, the remaining four users can also recover the desired sum as follows:
\begin{align}
&\trm{User  2:}~ X_1+X_3+X_4+X_5-\underbrace{{\bf H}_{\{1,2\}}S_{\{1,2\}}+{\bf H}_{\{2,3\}}S_{\{2,3\}}}_{Z_2}\notag\\
&+\underbrace{{\bf H}_{\{2,4\}}S_{\{2,4\}}+{\bf H}_{\{2,5\}}S_{\{2,5\}}}_{Z_2}  = W_1 + W_3+W_4+W_5,  \notag\\
&\trm{User  3:}~ X_1+X_2+X_4+X_5-\underbrace{{\bf H}_{\{1,3\}}S_{\{1,3\}}-{\bf H}_{\{2,3\}}S_{\{2,3\}}}_{Z_3}\notag\\
&+\underbrace{{\bf H}_{\{3,4\}}S_{\{3,4\}}+{\bf H}_{\{3,5\}}S_{\{3,5\}}}_{Z_3}  = W_1 + W_2+W_4+W_5,\notag\\
&\trm{User  4:}~ X_1+X_2+X_3+X_5-\underbrace{{\bf H}_{\{1,4\}}S_{\{1,4\}}-{\bf H}_{\{2,4\}}S_{\{2,4\}}}_{Z_4}\notag\\
&-\underbrace{{\bf H}_{\{3,4\}}S_{\{3,4\}}+{\bf H}_{\{4,5\}}S_{\{4,5\}}}_{Z_4}  = W_1+W_2+W_3+W_5,\notag\\
&\trm{User  5:}~ X_1+X_2+X_3+X_4-\underbrace{{\bf H}_{\{1,5\}}S_{\{1,5\}}-{\bf H}_{\{2,5\}}S_{\{2,5\}}}_{Z_5}\notag\\
&-\underbrace{{\bf H}_{\{3,5\}}S_{\{3,5\}}-{\bf H}_{\{4,5\}}S_{\{4,5\}}}_{Z_5}  = W_1+W_2+W_3+W_4.
\end{align}

We now show that  the proposed scheme satisfied the security constraint (\ref{eq:security constraint}). \Wlog, let us consider User 1 collude with User 2. By (\ref{eq:security constraint}), we have
\begin{align}
& I\left(X_2,X_3,X_4,X_5;W_2,W_3,W_4,W_5|\right.\notag\\
&\left.W_2+W_3+W_4+W_5, Z_1,W_1,Z_2,W_2\right)\notag\\
 =&H(X_2,X_3,X_4,X_5|W_2+W_3+W_4+W_5, Z_1,W_1,Z_2,W_2) \notag\\
&- H(X_2,X_3,X_4,X_5|W_3,W_4,W_5, Z_1,W_1,Z_2,W_2).\\
\leq &6-6=0. \label{eq:proof of security, example}
\end{align}
(\ref{eq:proof of security, example}) holds, as can be verified by the following proof. Specifically, we evaluate the two terms in (\ref{eq:proof of security, example}) separately.
First, the first term can be bounded as follows.  
\begin{align}
& H(X_2,X_3,X_4,X_5|W_2+W_3+W_4+W_5, Z_1,W_1,Z_2,W_2)\notag~~~~~~ \\
=& H(X_3,X_4,X_5|W_3+W_4+W_5, Z_1,W_1,Z_2,W_2) \\
=&  H(W_3-{\bf H}_{\{1,3\}}S_{\{1,3\}}-{\bf H}_{\{2,3\}}S_{\{2,3\}}+{\bf H}_{\{3,4\}}S_{\{3,4\}} \notag\\
&+{\bf H}_{\{3,5\}}S_{\{3,5\}},W_4-{\bf H}_{\{1,4\}}S_{\{1,4\}}-{\bf H}_{\{2,4\}}S_{\{2,4\}}\notag\\
&-{\bf H}_{\{3,4\}}S_{\{3,4\}}+{\bf H}_{\{4,5\}}S_{\{4,5\}},W_5-{\bf H}_{\{1,5\}}S_{\{1,5\}}\notag\\
&-{\bf H}_{\{2,5\}}S_{\{2,5\}}-{\bf H}_{\{3,5\}}S_{\{3,5\}}-{\bf H}_{\{4,5\}}S_{\{4,5\}}|\notag\\
&W_3+W_4+W_5,W_1,W_2,\notag\\
& S_{\{1,2\}}, S_{\{1,3\}}, S_{\{1,4\}},S_{\{1,5\}}, S_{\{2,3\}}, S_{\{2,4\}}, S_{\{2,5\}})\\
=&  H(W_3+{\bf H}_{\{3,4\}}S_{\{3,4\}} +{\bf H}_{\{3,5\}}S_{\{3,5\}},\notag\\
&W_4-{\bf H}_{\{3,4\}}S_{\{3,4\}}+{\bf H}_{\{4,5\}}S_{\{4,5\}},W_5\notag\\
&-{\bf H}_{\{3,5\}}S_{\{3,5\}}-{\bf H}_{\{4,5\}}S_{\{4,5\}}|W_3+W_4+W_5)\label{eq:ex2pfpf1}\\
=&  H(W_3+{\bf H}_{\{3,4\}}S_{\{3,4\}} +{\bf H}_{\{3,5\}}S_{\{3,5\}},\notag\\
&W_4-{\bf H}_{\{3,4\}}S_{\{3,4\}}+{\bf H}_{\{4,5\}}S_{\{4,5\}},W_5\notag\\
&-{\bf H}_{\{3,5\}}S_{\{3,5\}}-{\bf H}_{\{4,5\}}S_{\{4,5\}},W_3+W_4+W_5)\notag\\
&-H(W_3+W_4+W_5)\\
=&  H(W_3+{\bf H}_{\{3,4\}}S_{\{3,4\}} +{\bf H}_{\{3,5\}}S_{\{3,5\}},\notag\\
&W_4-{\bf H}_{\{3,4\}}S_{\{3,4\}}+{\bf H}_{\{4,5\}}S_{\{4,5\}},W_5-{\bf H}_{\{3,5\}}S_{\{3,5\}}\notag\\
&-{\bf H}_{\{4,5\}}S_{\{4,5\}})-H(W_3+W_4+W_5)\label{eq:ex2pfpf2}\\
\leq&  H(W_3+{\bf H}_{\{3,4\}}S_{\{3,4\}} +{\bf H}_{\{3,5\}}S_{\{3,5\}})+\notag\\
&H(W_4-{\bf H}_{\{3,4\}}S_{\{3,4\}}+{\bf H}_{\{4,5\}}S_{\{4,5\}})+\notag\\
&H(W_5-{\bf H}_{\{3,5\}}S_{\{3,5\}})-{\bf H}_{\{4,5\}}S_{\{4,5\}})\notag\\
&-H(W_3+W_4+W_5)\\
=&3L-L=2L=6,\label{eq:step1,term 2, proof of security, example}
\end{align}
where (\ref{eq:ex2pfpf1}) is due to the independence of the keys and inputs. 
(\ref{eq:ex2pfpf2}) is due to $W_3+W_4+W_5$ can be derived from $W_3+{\bf H}_{\{3,4\}}S_{\{3,4\}} +{\bf H}_{\{3,5\}}S_{\{3,5\}},$ $W_4-{\bf H}_{\{3,4\}}S_{\{3,4\}}+{\bf H}_{\{4,5\}}S_{\{4,5\}},$ and $W_5-{\bf H}_{\{3,5\}}S_{\{3,5\}})-{\bf H}_{\{4,5\}}S_{\{4,5\}}$.
Second,  the second term in (\ref{eq:proof of security, example}) is equal to
\begin{align}
& H(X_2,X_3,X_4,X_5|W_3,W_4,W_5, Z_1,W_1,Z_2,W_2)\notag \\
=&  H\left( W_2-{\bf H}_{\{1,2\}}S_{\{1,2\}}+{\bf H}_{\{2,3\}}S_{\{2,3\}}+{\bf H}_{\{2,4\}}S_{\{2,4\}}\right.  \notag\\
&+{\bf H}_{\{2,5\}}S_{\{2,5\}},W_3-{\bf H}_{\{1,3\}}S_{\{1,3\}}-{\bf H}_{\{2,3\}}S_{\{2,3\}} \notag\\
&+{\bf H}_{\{3,4\}}S_{\{3,4\}}+{\bf H}_{\{3,5\}}S_{\{3,5\}},W_4-{\bf H}_{\{1,4\}}S_{\{1,4\}}\notag\\
&-{\bf H}_{\{2,4\}}S_{\{2,4\}}-{\bf H}_{\{3,4\}}S_{\{3,4\}}+{\bf H}_{\{4,5\}}S_{\{4,5\}},\notag\\
&W_5-{\bf H}_{\{1,5\}}S_{\{1,5\}}-{\bf H}_{\{2,5\}}S_{\{2,5\}}-{\bf H}_{\{3,5\}}S_{\{3,5\}}\notag\\
&-{\bf H}_{\{4,5\}}S_{\{4,5\}}|W_2,W_3,W_4,W_5, W_1,S_{\{1,2\}}, S_{\{1,3\}},\notag\\
& S_{\{1,4\}}, S_{\{1,5\}}, S_{\{2,3\}}, S_{\{2,4\}}, S_{\{2,5\}})\\
=&  H\left({\bf H}_{\{3,4\}}S_{\{3,4\}}+{\bf H}_{\{3,5\}}S_{\{3,5\}},-{\bf H}_{\{3,4\}}S_{\{3,4\}}\right.\notag\\
&+{\bf H}_{\{4,5\}}S_{\{4,5\}},-{\bf H}_{\{3,5\}}S_{\{3,5\}}-{\bf H}_{\{4,5\}}S_{\{4,5\}}| \notag\\
&W_2,W_3,W_4,W_5, W_1,S_{\{1,2\}}, S_{\{1,3\}}, S_{\{1,4\}}, S_{\{1,5\}},\notag\\
& S_{\{2,3\}}, S_{\{2,4\}}, S_{\{2,5\}})\\ 
=&  H\left({\bf H}_{\{3,4\}}S_{\{3,4\}}+{\bf H}_{\{3,5\}}S_{\{3,5\}},-{\bf H}_{\{3,4\}}S_{\{3,4\}}\right.\notag\\
&+{\bf H}_{\{4,5\}}S_{\{4,5\}},-{\bf H}_{\{3,5\}}S_{\{3,5\}}-{\bf H}_{\{4,5\}}S_{\{4,5\}} )\label{eq:ex2pf1} \\ 
=&  H\left( S_{\{3,4\}},S_{\{3,5\}},S_{\{4,5\}}\right)\label{eq:ex2pf2} \\
=&3L_S=2L=6,
\end{align}
where (\ref{eq:ex2pf1}) is due to the \indepce of the keys and inputs. (\ref{eq:ex2pf2}) holds because the three vectors 
$ {\bf H}_{\{3,4\}}S_{\{3,4\}}+{\bf H}_{\{3,5\}}S_{\{3,5\}},$ $ 
-{\bf H}_{\{3,4\}}S_{\{3,4\}}+{\bf H}_{\{4,5\}}S_{\{4,5\}},$ $
-{\bf H}_{\{3,5\}}S_{\{3,5\}}-{\bf H}_{\{4,5\}}S_{\{4,5\}} $ 
are linearly independent, and their construction is carefully designed as shown in (\ref{V_example2}).

Since \muinfo is non-negative, we conclude that $I(X_2,X_3,X_4,X_5;W_2,W_3,W_4,W_5|W_2+W_3+W_4+W_5,$ $ Z_1,W_1,Z_2,W_2)=0$, proving the security for User 1. 
The proofs for other  users follow similarly. 
\hfill $ \lozenge$
\end{example}

\subsection{Achievable Scheme for Arbitrary $K,T$ and $G$}
\label{subsec: general scheme}
Following the above examples, we now describe the general achievable scheme. 
To achieve the communication rate $R_X=1$ and the groupwise rate $R_S=(K-T-2)/\binom{K-T-1}{G}$, we set each input to consist of $L=K!\binom{K-T-1}{G}$ symbols over the finite field $\mbb{F}_q$, i.e., $W_k\in \mbb{F}_q^{L\times 1}$, and let $L_X = L$. 
Similarly, for each group $\mathcal{G}\subseteq [K]$ with $1<|\mathcal{G}|=G\leq K-T-1$, we assign a groupwise key consisting of $L_S=K!(K-T-2)$ symbols, i.e., $S_\mathcal{G}\in \mbb{F}_q^{L_S\times 1}$. 
For any user $k\in [K]$, we set the message as
\begin{eqnarray}
    X_k &=& W_k + \sum_{{\mathcal{G}:k \in \mathcal{G},\mathcal{G}\in \binom{[K]}{G}}} {\bf H}_{\mathcal{G}}^{k}S_{\mathcal{G}},~\forall k\in[K], \label{eq:sc11}
\end{eqnarray}
where ${\bf H}_{\mathcal{G}}^{k}$ is an {\color{black}$L\times L_S$} matrix over $\mathbb{F}_{q}$, serving as the coefficient of the groupwise key. 
To ensure both the correctness constraint (\ref{eq:recovery constraint}) and the security constraint (\ref{eq:security constraint}), the matrices ${\bf H}_{\mathcal{G}}^{k}$ must be carefully designed for all $k\in [K]$ and $\mathcal{G}\subset [K]$. Next, we demonstrate the existence of such matrices ${\bf H}_{\mathcal{G}}^{k},k\in [K],\mathcal{G}\subset [K]$.

To ensure correctness, the sum of the groupwise keys must equal zero. To satisfy this constraint, we set
\begin{eqnarray}\label{Vsum1}
    \sum_{k \in \mathcal{G}}{\bf H}_{\mathcal{G}}^k={\bf 0}_{L\times L_S},~\forall \mathcal{G}\in \binom{[K]}{G} \label{eq:zerosum1}
\end{eqnarray}
such that 
\begin{eqnarray}\label{Vsum2}
    \left(\sum_{k \in \mathcal{G}}{\bf H}_{\mathcal{G}}^k \right)S_{\mathcal{G}}={\bf 0}_{L\times 1},~\forall \mathcal{G}\in \binom{[K]}{G}, \label{eq:zerosum2}
\end{eqnarray}
which ensure that for any  $\mathcal{G}\in \binom{[K]}{G}$, the sum of the groupwise key is zero.
In words, the groupwise-key contribution vanishes in the sum for any $\mathcal{G}$.

Let us check the correctness constraint (\ref{eq:recovery constraint}), for any user $u\in [K]$, $X_u$ can be derived from $W_u$ and $Z_u$, we have 
\begin{align}
     \sum_{k \in [K]} X_k =& \sum_{k \in [K]} W_k + \sum_{k\in[K]} \sum_{{\mathcal{G}:k \in \mathcal{G},\mathcal{G}\in \binom{[K]}{G}}} {\bf H}_{\mathcal{G}}^{k}S_{\mathcal{G}}\\ 
    =& \sum_{k \in [K]}W_k +\underbrace{\sum_{\mathcal{G}:k\in\mathcal{G},\mathcal{G}\in\binom{[K]}{G}} \left[\left(\sum_{k \in \mathcal{G}}{\bf H}_{\mathcal{G}}^k\right)S_{\mathcal{G}}\right]}_{\overset{(\ref{eq:zerosum2})}{=}0}\label{eq:crnspf1}\\
    =& \sum_{k \in [K]} W_k.
\end{align}
The second term in (\ref{eq:crnspf1}) is zero because ${\bf H}_{\mathcal{G}}^{k}$ is designed so that (\ref{Vsum2}) holds. The design of ${\bf H}_{\mathcal{G}}^{k}$ for the  correctness constraint is complete, we now consider the design of ${\bf H}_{\mathcal{G}}^{k}$ for the security constraint.

Consider the security constraint (\ref{eq:security constraint}), for any user $k\in [K]$, we have
\begin{align}
    0=&I\left(\left\{W_i\right\}_{i\in[K]\setminus \{k\}}; \left\{X_i\right\}_{i\in[K]\setminus\{k\}} \Bigg|\right.\notag\\
    &\left.\sum_{i\in [K]} W_i, \left\{ W_i, Z_i \right\}_{i\in\mathcal{T}},Z_k,W_k \right)\notag\\
    =&I\left(\left\{W_i\right\}_{i\in[K]\setminus (\mathcal{T}\cup\{k\})}; \left\{X_i\right\}_{i\in[K]\setminus(\mathcal{T}\cup\{k\})} \Bigg|\right.\notag\\
    &\left.\sum_{i\in [K]} W_i,\left\{ W_i, Z_i \right\}_{i\in\mathcal{T}},Z_k,W_k \right)\notag\\
    =&H\left(\left\{X_i\right\}_{i\in[K]\setminus(\mathcal{T}\cup\{k\})} \Bigg|\sum_{i\in [K]} W_i, \left\{ W_i, Z_i \right\}_{i\in(\mathcal{T}\cup\{k\})} \right)\notag\\
    &-H\left(\left\{X_i\right\}_{i\in[K]\setminus(\mathcal{T}\cup\{k\})} \Bigg|\sum_{i\in [K]} W_i,\right. \notag\\
    &\left.\left\{ W_i, Z_i \right\}_{i\in(\mathcal{T}\cup\{k\})},\left\{W_i\right\}_{i\in[K]\setminus(\mathcal{T}\cup\{k\})} \right)\notag\\
    \leq &(|[K]\setminus(\mathcal{T}\cup\{k\})|-1)L\notag\\
    & ~- H\left(\left\{
    \sum_{{\mathcal{G}:i \in \mathcal{G},\mathcal{G}\in \binom{[K]\backslash (\mathcal{T}\cup\{k\})}{G}}} {\bf H}_{\mathcal{G}}^{i}S_{\mathcal{G}}
    \right\}_{i\in[K]\backslash(\mathcal{T}\cup\{k\})}\Bigg|\right.\notag\\
    &\left.\{W_i\}_{i\in[K]},\left\{ Z_i \right\}_{i\in(\mathcal{T}\cup\{k\})}\right)\label{lemma:rank_eq211}\\
    \overset{(\ref{ind_S})}{=}& (K-|\mathcal{T}|-2)L \notag\\
    & ~- H\left(\left\{
    \sum_{{\mathcal{G}:i \in \mathcal{G},\mathcal{G}\in \binom{[K]\backslash (\mathcal{T}\cup\{k\})}{G}}} {\bf H}_{\mathcal{G}}^{i}S_{\mathcal{G}}
    \right\}_{i\in[K]\backslash(\mathcal{T}\cup\{k\})}\right),\label{eq:ss11}
\end{align}
where in (\ref{lemma:rank_eq211}), the first term follows from the fact that $\sum_{i\in[K]\backslash(\mathcal{T}\cup\{k\})} X_i = \sum_{i\in[K]\backslash(\mathcal{T}\cup\{k\})} W_i$ and uniform distribution maximizes entropy, while the second term holds because $\big\{
\sum_{{\mathcal{G}:i \in \mathcal{G},\mathcal{G}\in \binom{[K]\backslash (\mathcal{T}\cup\{k\})}{G}}} {\bf H}_{\mathcal{G}}^{i}S_{\mathcal{G}}
\big\}_{i\in[K]\backslash(\mathcal{T}\cup\{k\})}$ only involves groupwise keys that are unknown to user $k$ and the colluding set $\mathcal{T}$.
The second term of (\ref{eq:ss11}) holds because the groupwise keys $\big\{
\sum_{\substack{\mathcal{G}\in \binom{[K]\setminus (\mathcal{T}\cup\{k\})}{G},\, i \in \mathcal{G}}} {\bf H}_{\mathcal{G}}^{i}S_{\mathcal{G}}
\big\}_{i\in[K]\setminus(\mathcal{T}\cup\{k\})}$
are independent of the inputs $\{W_i\}_{i\in [K]}$ and exclude the groupwise keys associated with user $k$ and the colluding set $\mathcal{T}$, i.e., $\{Z_i\}_{i\in(\mathcal{T}\cup\{k\})}$.
Hence, by (\ref{eq:ss11}), the security constraint is satisfied if
\begin{align}
&H\left(\left\{
\sum_{{\mathcal{G}:i \in \mathcal{G},\mathcal{G}\in \binom{[K]\backslash (\mathcal{T}\cup\{k\})}{G}}} {\bf H}_{\mathcal{G}}^{i}S_{\mathcal{G}}
    \right\}_{i\in[K]\backslash(\mathcal{T}\cup\{k\})}\right)\notag\\
& \geq  (K-|\mathcal{T}|-2)L.\label{eq:matrsec}
\end{align}

Next, we demonstrate the existence of matrices ${\bf H}_{\mathcal{G}}^{i}$ such that (\ref{eq:matrsec}) holds.
Collecting all messages and writing them in a matrix form, we have
\begin{eqnarray}
\left[
    \begin{array}{c}
         X_1\\
         \vdots\\
         X_K
    \end{array}\right]=\left[
    \begin{array}{c}
         W_1\\
         \vdots\\
         W_K
    \end{array}\right]+ {\bf H}
    \left[\begin{array}{c}
         S_{\{1,2,\cdots,G\}}\\
         \vdots\\
         S_{\{K-G+1,\cdots,K\}}
    \end{array}
    \right], \label{eq:messX}
\end{eqnarray}
where
\begin{align}
    {\bf H} =& \left[ \left({\bf H}_{\mathcal{G}}^k\right)_{k\in[K], \mathcal{G} \in \binom{[K]}{G}} \right]\notag\\
    \eqdef& \left[\begin{array}{cccc}
        {\bf H}_{\{1,2,\cdots,G\}}^1  &\cdots& {\bf H}_{\{K-G+1,\cdots,K\}}^1\\
        \vdots  & \ddots &\vdots\\
        {\bf H}_{\{1,2,\cdots,G\}}^K &\cdots& {\bf H}_{\{K-G+1,\cdots,K\}}^K\\
    \end{array}\right], \label{eq:h}
\end{align}
and ${\bf H}_{\mathcal{G}}^k={\bf 0}_{L\times L_S}$, 
$\forall k \in [K], \mathcal{G} \in \binom{[K]}{G}, k \notin \mathcal{G}$.
We denote by $\hat{\mathbf{H}}[\mathcal{T}\cup{k}]$ the submatrix of $\mathbf{H}$ containing the coefficients corresponding to the groupwise keys that are unknown to the colluding set $\mathcal{T}$ and user $k$, i.e.,
\begin{eqnarray}
\hat{\bf H}[\mathcal{T}\cup\{k\}] \triangleq 
\left[ 
    \left({\bf H}_{\mathcal{G}}^i\right)_{i\in [K]\setminus(\mathcal{T}\cup\{k\}),\, 
    \mathcal{G}\in \binom{[K]\setminus(\mathcal{T}\cup\{k\})}{G}} 
\right],
\label{eq:hhat}
\end{eqnarray}
which contains $K-|\mathcal{T}|-1$ row blocks and $\binom{K-|\mathcal{T}|-1}{G}$ column blocks of ${\bf H}_{\mathcal{G}}^i$ terms. 
Then $\big\{\sum_{{\mathcal{G}:i \in \mathcal{G},\mathcal{G}\in \binom{[K]\backslash (\mathcal{T}\cup\{k\})}{G}}} {\bf H}_{\mathcal{G}}^{i}S_{\mathcal{G}}
\big\}_{i\in[K]\backslash(\mathcal{T}\cup\{k\})}$ can be written as 
\begin{align}
    &\left\{
    \sum_{{\mathcal{G}:i \in \mathcal{G},\mathcal{G}\in \binom{[K]\backslash (\mathcal{T}\cup\{k\})}{G}}} {\bf H}_{\mathcal{G}}^{i}S_{\mathcal{G}}\right\}_{i\in[K]\backslash(\mathcal{T}\cup\{k\})}\notag\\
    & = \hat{\bf H}[\mathcal{T}\cup\{k\}] \left(S_{\mathcal{G}}\right)_{\mathcal{G}\in \binom{[K]\backslash (\mathcal{T}\cup\{k\})}{G}}.
\end{align}

Next, we show that there exists a matrix ${\bf H}$ of form (\ref{eq:h}) such that its submatrix $\hat{\bf H}[\mathcal{T}\cup\{k\}]$, specified in (\ref{eq:hhat}), has rank $(K-|\mathcal{T}|-2)L$ over $\mathbb{F}_{q}$ for all possible colluding user sets $\mathcal{T}, \forall \mathcal{T} \subset [K]\backslash\{k\}, |\mathcal{T}| \leq T$.

We show that if for each $\mathcal{G} \in \binom{[K]}{G}$, we generate each element of any $G-1$ matrices ${\bf H}_\mathcal{G}^i, i \in \mathcal{G}$ in (\ref{eq:h}) uniformly and i.i.d. over the field $\mathbb{F}_{q}$ (and the last matrix is set as the negative of the sum of the remaining $G-1$ matrices, to satisfy (\ref{eq:zerosum1})), then as
$q \rightarrow \infty$, the probability that $\hat{\bf H}[\mathcal{T}\cup\{k\}]$ has rank $(K-|\mathcal{T}|-2)L$ approaches $1$ such that the existence of 
${\bf H}$ is guaranteed. 
To apply the Schwartz-Zippel lemma, 
we need to guarantee that for each $\mathcal{T}$, there exists a realization of $\hat{\bf H}[\mathcal{T}\cup\{k\}]$ such that $\mbox{rank}(\hat{\bf H}[\mathcal{T}\cup\{k\}]) = (K-|\mathcal{T}|-2)L$.

We are left to show that for any fixed $\mathcal{T}$, we have an $\hat{\bf H}[\mathcal{T}\cup\{k\}]$ of rank $(K-|\mathcal{T}|-2)L$ over $\mathbb{F}_{q}$.
Consider any $\mathcal{T}\subset [K]$ and $k\in [K]\setminus \mathcal{T}$.
Note that $\hat{\bf H}[\mathcal{T}\cup\{k\}]$ is the same as the matrix $\bf H$ in (\ref{eq:h}) when we have users $K-|\mathcal{T}|-1$ in $[K]\setminus (\mathcal{T}\cap \{k\})$, $0$ colluding users, and groupwise keys of group size $G$.

Denote the new matrix as $\mathbf H'$, which is of size $(K-|\mathcal T|-1)L \times \binom{K-|\mathcal T|-1}{G}L_S$.
We construct $\mathbf H'$ to have rank $(K-|\mathcal T|-2)L$. Starting from a base scheme of length $L' = 1$, for each user $i \in [K]\setminus(\mathcal T\cup{k})$, define the messages as the form in (\ref{eq:sc11}). 
By applying all possible permutations of the user indices and repeating this basic construction, we obtain a scheme of length $L' = K!$ with symmetric groupwise keys of length $L_{S}' = K!(K-T-2)/\binom{K-T-1}{G}$. For each repetition, we use independent inputs and groupwise keys.
Then, the messages of users in \([K]\setminus(\mathcal{T}\cup\{k\})\) are constructed as in (\ref{eq:messX}).
Due to the repetition of basic scheme and the independence and uniformity of inputs and groupwise key, matrix ${\bf H'}$ has rank $L'(K-T-2)=K!(K-T-2)$. Then, by repeating the scheme \(\binom{K - T - 1}{G}\) times, we obtain the desired input length 
\(L = L'\binom{K - T - 1}{G} = K!\binom{K - T - 1}{G}\) 
and a matrix \(\mathbf{H}'\) with the desired rank 
\((K - T - 2)K!\binom{K - T - 1}{G}\), 
which satisfy the constraint in (\ref{eq:matrsec}).

\section{Converse}
\label{sec: converse}
This section presents the converse proof which establishes lower bounds on the \comm and \grpws key rates.
To illustrate the main idea, we first revisit Example 2 and then present the general converse proof.

\subsection{Converse Proof of Example 2}

For Example 2, we show that any secure aggregation scheme must satisfy the following bounds:
\begin{eqnarray}
\label{eq:lower bounds,converse,example}
R_X \ge 1, \quad R_S \ge 2/3.
\end{eqnarray}

We first establish the following intermediate result, which asserts that the entropy of $X_1$, conditioned on the inputs and keys of the other users, is lower bounded by $L$ symbols:
\begin{eqnarray}
    \label{eq:result1,H(X1|(Wi,Zi)i=2,3)>=L,example}
H\left(X_1|W_2,W_3,W_4,W_5, Z_2,Z_3, Z_4,Z_5\right) \geq L.
\end{eqnarray}
Here is the intuitive explanation. Since any other user needs to compute the sum of all inputs excluding its own, the sum necessarily involves $W_1$. As $W_1$ is available only at User 1, it must be conveyed through $X_1$. Therefore, the entropy of $X_1$ is at least that of $W_1$, i.e., $L$.
More formally, we have
\begin{align}
& H\left(X_1| W_2,W_3, W_4,W_5,Z_2,Z_3 ,Z_4,Z_5  \right) \notag \\ 
  \ge & I\left(X_1; W_1|\{W_i,Z_i\}_{i \in\{2,3,4,5\} }\right)\\
 = & H\left(W_1|\{W_i,Z_i\}_{i \in\{2,3,4,5\} }\right)\notag\\
&  - H\left(W_1|\{W_i,Z_i\}_{i \in\{2,3,4,5\}},X_1\right) \\
 = & H\left(W_1\right) - H\left(W_1|\{W_i,Z_i,X_i\}_{i \in\{2,3,4,5\}},\right.\notag\\
&\left. X_1,W_1+W_2+W_3+W_4+W_5\right)
\label{eq:step 0, proof of lemma, example}\\
 = &  H(W_1)\label{eq:step 1, proof of lemma, example}\\
=&L.
\end{align}
In (\ref{eq:step 0, proof of lemma, example}), the first term follows from the independence of the inputs and the keys. The second term holds because $X_2,X_3,X_4,X_5$ are deterministic functions of $(W_2,Z_2),(W_3,Z_3),(W_4,Z_4),(W_5,Z_5)$, while the global sum $W_1+W_2+W_3+W_4+W_5$ can be obtained from $(X_1,X_2,X_3,X_4,X_5)$. Furthermore, (\ref{eq:step 1, proof of lemma, example}) holds since $W_1$ is recoverable from $(W_2,W_3,W_4,W_5)$ and the global sum $W_1+W_2+W_3+W_4+W_5$.
Similarly, we can also obtain
\be
\label{eq:result1,H(X2|(Wi,Zi)i=1,3)>=L,example}
H\left(X_k|\{W_i,Z_i\}_{i\in [5]\setminus\{k\}}\right) \ge L, \forall k\in[5].
\ee 

\tit{1) Proof  of $R_X\geq 1$.}
Equipped with  (\ref{eq:result1,H(X1|(Wi,Zi)i=2,3)>=L,example}), $R_X\geq 1$ follows immediately: 
\begin{align}
 H(X_1) & \ge H(X_1|W_2,W_3, W_4,W_5,Z_2,Z_3 ,Z_4,Z_5) 
 \overset{(\ref{eq:result1,H(X1|(Wi,Zi)i=2,3)>=L,example})}{\ge} L\notag\\
& \Rightarrow R_X \eqdef \frac{L_X}{L}  \ge \frac{H(X_1)}{L}\ge 1. 
\end{align}
The condition $\rx \ge 1$ is natural, since each transmitted message $X_k$ is required to contain the complete information of the input $W_k$, leading to $L_X \ge L$ for every $k \in [5]$.

Next, we derive the lower bound of the groupwise key rate for Example 2. We first prove a useful result. For the security constraint, the message $X_1$ should not reveal any information about the input $W_1$; in other words, the mutual information between $X_1$ and $W_1$ is zero, even when conditioned on the information of one of the other users.
\begin{eqnarray}
    \label{eq:I(X1;W1|Wk,Zk)=0,indetermediate result,example}
I\left( X_1; W_1 | W_k, Z_k\right)=0,\; \forall k \in\{2,3,4,5\}.
\end{eqnarray}

\Wlog, let us consider $k=2$:
\begin{align}
& I\left( X_1; W_1 | W_2, Z_2\right)\notag \\
 \leq& I\left( X_1,W_1+W_2+W_3+W_4+W_5; W_1 | W_2, Z_2\right)\\
  =& I\left( W_1+W_2+W_3+W_4+W_5; W_1| W_2, Z_2\right)\notag\\
&   +  I\left( X_1; W_1|W_1+W_2+W_3+W_4+W_5, W_2, Z_2\right)\\
 \leq& \underbrace{ I\left( W_1+W_3+W_4+W_5; W_1| W_2, Z_2\right)}_{\overset{(\ref{ind_S})}{=}0}\notag\\
&   + \underbrace{I( X_1,X_3,X_4,X_5 ; W_1,W_3,W_4,W_5|}_{\ovst{(\ref{eq:security constraint})}{=}0  }\notag\\
&\underbrace{W_1+W_2+W_3+W_4+W_5, W_2, Z_2)}_{\ovst{(\ref{eq:security constraint})}{=}0  }\label{eq:step0,proof,I(X1;W1|Wk,Zk)=0,indetermediate result,example}\\
 = & 0,
\end{align}
where in (\ref{eq:step0,proof,I(X1;W1|Wk,Zk)=0,indetermediate result,example}), the first term is zero due to the independence of the inputs $W_{1:K}$ from each other and from the keys. The second term is zero as a direct consequence of the security requirement for user 2.

Second, we show that the joint entropy of the messages $X_1, X_2, X_3$, conditioned on the information available to Users 4 and 5, is lower bounded by $3L$. 
\begin{eqnarray}
     H(X_1, X_2,X_3|W_4, Z_4,W_5,Z_5) \geq 3L. \label{eq:proof H(X1,X2|W3,Z3)>=2L,example1}
\end{eqnarray}
The intuition is that each of the inputs $W_1, W_2$, and $W_3$ must be fully contained in the corresponding messages $X_1, X_2$, and $X_3$, respectively. Consequently, the joint entropy of $X_1, X_2, X_3$ cannot be less than $3L$. A rigorous proof is provided below.
\begin{align}
& H(X_1, X_2,X_3|W_4, Z_4,W_5,Z_5) \notag\\
 =& H(X_1|W_4, Z_4,W_5,Z_5) + H(X_2|W_4, Z_4,W_5,Z_5, X_1) \notag\\
&+H(X_3|W_4, Z_4,W_5,Z_5, X_1,X_2) \\
  \geq & H(X_1|W_2,Z_2,W_3,Z_3,W_4, Z_4,W_5,Z_5) \notag\\
&+ H(X_2|W_3,Z_3,W_4, Z_4,W_5,Z_5,W_1,Z_1, X_1) \notag\\
&+H(X_3|W_4, Z_4,W_5,Z_5,W_1,Z_1,W_2,Z_2, X_1,X_2) \\
= & H(X_1|W_2,Z_2,W_3,Z_3,W_4, Z_4,W_5,Z_5) \notag\\
&+ H(X_2|W_3,Z_3,W_4, Z_4,W_5,Z_5,W_1,Z_1) \notag\\
&+H(X_3|W_4, Z_4,W_5,Z_5,W_1,Z_1,W_2,Z_2) \\
 & \ovst{(\ref{eq:result1,H(X1|(Wi,Zi)i=2,3)>=L,example})(\ref{eq:result1,H(X2|(Wi,Zi)i=1,3)>=L,example})}{\ge } 3L.\label{eq:step0,proof H(X1,X2|W3,Z3)>=2L,example}
\end{align}

Third, we establish that the mutual information between $(X_1,X_2,X_3)$ and $(W_1,W_2,W_3)$, conditioned on $(W_4,Z_4,W_5,Z_5)$, is $L$. The reasoning is that, from the viewpoint of user 4 colluding with user 5, the only recoverable information from the set of messages $(X_1,X_2,X_3)$ is the sum $W_1+W_2+W_3$, which, owing to the uniform distribution of the inputs, contains exactly $L$ symbols. In particular, we have
\begin{eqnarray}
    I(X_1, X_2,X_3;W_1, W_2,W_3|W_4, Z_4,W_5,Z_5)=L.\label{eq:proof,I(X_1,X_2;W_1,W_2|W_3,Z_3)=L}
\end{eqnarray}
Here is the detailed proof.
\begin{align}
&I(X_1, X_2,X_3;W_1, W_2,W_3|W_4, Z_4,W_5,Z_5) \notag\\
 =& I(X_1, X_2,X_3;W_1, W_2,W_3,W_1+ W_2+W_3|\notag\\
&W_4, Z_4,W_5,Z_5)\\
 =& I(X_1, X_2,X_3;W_1+ W_2+W_3|W_4, Z_4,W_5,Z_5)\notag\\
  &+\underbrace{I(X_1, X_2,X_3;W_1, W_2,W_3|}_{\overset{(\ref{eq:security constraint})}{=}0}\notag\\
  &\underbrace{W_1+ W_2+W_3,W_4, Z_4,W_5,Z_5)}_{\overset{(\ref{eq:security constraint})}{=}0}\label{convpf1}\\
   =& H(W_1+ W_2+W_3|W_4, Z_4,W_5,Z_5)\notag\\
    & -\underbrace{H(W_1+ W_2+W_3|X_1, X_2,X_3,W_4, Z_4,W_5,Z_5)}_{\overset{(\ref{eq:recovery constraint})}{=}0}\label{convpf2}\\
  \ovst{(\ref{eq:input independence})}{=}& H(W_1+W_2+W_3)=L.
\end{align}
Specifically, the second term in (\ref{convpf1}) is zero by the security constraint (\ref{eq:security constraint}), and the second term in (\ref{convpf2}) is zero by the correctness constraint (\ref{eq:recovery constraint}). The last two steps rely on the independence between the keys and inputs together with the uniformity of the inputs.

Fourth, the lower bound of joint entropy of $Z_1,Z_2$ and $Z_3$ is $2L$, \ie, 
\begin{eqnarray}
    H(Z_1,Z_2,Z_3|Z_4,Z_5)\geq 2L,\label{eq:proof,H(Z1,Z2|Z3)>=L,example}
\end{eqnarray}
which can be proved \af.
\begin{align}
& H(Z_1,Z_2,Z_3|Z_4,Z_5) \notag\\
\ovst{(\ref{ind_S})}{=}&   H(Z_1,Z_2,Z_3|W_1,W_2,W_3,W_4,W_5,Z_4,Z_5) \label{convsepf1}\\
\geq &I(Z_1,Z_2,Z_3;X_1,X_2,X_3|\{W_i\}_{i\in [5]},Z_4,Z_5) \label{convsepf2}\\
=&H(X_1,X_2,X_3|\{W_i\}_{i\in [5]},Z_4,Z_5) \notag\\
&-\underbrace{H(X_1,X_2,X_3|\{W_i\}_{i\in [5]},\{Z_i\}_{i\in [5]})}_{\overset{(\ref{message})}{=}0} \\
=&H(X_1,X_2,X_3|W_4,W_5,Z_4,Z_5) \notag\\
&-I(X_1,X_2,X_3;W_1,W_2,W_3|W_4,W_5,Z_4,Z_5) \\
\overset{(\ref{eq:proof H(X1,X2|W3,Z3)>=2L,example1})(\ref{eq:proof,I(X_1,X_2;W_1,W_2|W_3,Z_3)=L})}{\geq}&3L-L=2L,
\end{align}
where (\ref{convsepf1}) is due to the \indepce  of the inputs and keys. In the last step, we applied the two previously obtained bounds (\ref{eq:proof H(X1,X2|W3,Z3)>=2L,example1}) and (\ref{eq:proof,I(X_1,X_2;W_1,W_2|W_3,Z_3)=L}).

\tit{2) Proof of $R_S \ge 2/3$.}
With the above intermediate results (\ref{eq:proof,H(Z1,Z2|Z3)>=L,example}), we are now ready to prove $R_S \geq 2/3$:
\begin{align}
2L \overset{(\ref{eq:proof,H(Z1,Z2|Z3)>=L,example})}{\leq}& H\left(Z_1,Z_2,Z_3|Z_4,Z_5\right)\\
\overset{(\ref{Z_individual})}{=}&H(S_{\{1,2\}},S_{\{1,3\}},S_{\{1,4\}},S_{\{1,5\}},S_{\{2,3\}},S_{\{2,4\}}, S_{\{2,5\}}, \notag\\
&S_{\{3,4\}},S_{\{3,5\}},S_{\{4,5\}}|S_{\{1,4\}},S_{\{1,5\}}S_{\{2,4\}}, S_{\{2,5\}},\notag\\
&S_{\{3,4\}},S_{\{3,5\}},S_{\{4,5\}})\\
=&H(S_{\{1,2\}},S_{\{1,3\}},S_{\{2,3\}}|S_{\{1,4\}},S_{\{1,5\}}S_{\{2,4\}}, S_{\{2,5\}},\notag\\
&S_{\{3,4\}},S_{\{3,5\}},S_{\{4,5\}})\\
=&H(S_{1,2},S_{\{1,3\}},S_{\{2,3\}})\\
\overset{(\ref{ind_S})}{=}&3L_S,\\
\Rightarrow~~~&R_S=\frac{L_S}{L}\geq \frac{2}{3}.
\end{align}

\subsection{General Converse Proof}
We are now ready to generalize the above proof to all parameter settings.
We begin by proving the infeasible regime.

\subsubsection{Infeasibility  at $G=1$ or $G\geq K-T$} \label{subsec:infeasibility proof, converse}
When $G=1$, the keys are mutually independent. Consider User 1 and User 2. Let $\mathcal{U}=[K]\setminus (\mathcal{T}\cup \{1,2\})$ with $|\mathcal{U}|\geq 1$. Then
\begin{eqnarray}
I\left(Z_1;\{Z_k\}_{k \in \mathcal{U}} \Big|Z_2,\{Z_k\}_{k \in \mathcal{T}}\right) \overset{(\ref{Z_individual})}{=} 0, \label{thm3_eq6}
\end{eqnarray}
since for $G=1$ the collection $\{Z_k\}_{k\in [K]}$ is mutually independent.
Next, we show that the above mutual information term must be strictly positive, i.e., the keys must be correlated.
\begin{align}
    &  I\left(Z_1;\{Z_k\}_{k \in \mathcal{U}} \Big|Z_2, \{Z_k\}_{k \in \mathcal{T}}\right) \notag\\
   \overset{(\ref{ind_S})}{=}& I\left(W_1,Z_1;\{W_k,Z_k\}_{k \in \mathcal{U}}|Z_2,W_2,\{W_k,Z_k\}_{k \in \mathcal{T}}\right)\label{thm3_eq1}\\
   \overset{(\ref{message})}{\geq}& I\left(W_1,X_1;\{W_k,X_k\}_{k \in \mathcal{U}}|Z_2,W_2,\{W_k,Z_k\}_{k \in \mathcal{T}}\right)\label{thm3_eq2}\\
   \geq& I\left( W_1;\{W_k,X_k\}_{k \in \mathcal{U}}\big|X_1,Z_2,W_2,\{W_k,Z_k\}_{k \in \mathcal{T}}\right)\\
   =& H\left( W_1 \big|X_1,Z_2,W_2,\{W_k,Z_k\}_{k \in \mathcal{T}}\right)-\notag\\
   &\underbrace{H\left(W_1 \big|X_1,Z_2,W_2,\{W_k,X_k\}_{k \in \mathcal{U}},\{W_k,Z_k\}_{k \in \mathcal{T}}\right)}_{\overset{(\ref{message})(\ref{eq:recovery constraint})}{=}0}\label{thm3_eq3}\\
   =& H\left( W_1 \big|Z_2,W_2,\{W_k,Z_k\}_{k \in \mathcal{T}}\right)\notag\\
   &-I\left(W_1;X_1 \big|Z_2,W_2,\{W_k,Z_k\}_{k \in \mathcal{T}}\right)\label{thm3_eq4}\\
   \geq& H\left( W_1 \big|Z_2,W_2,\{W_k,Z_k\}_{k \in \mathcal{T}}\right)\notag\\
   &- I\left(W_1;X_1, \sum_{i=1}^K W_i \Bigg|Z_2,W_2,\{W_k,Z_k\}_{k \in \mathcal{T}}\right)\\
   =& H\left( W_1\right)- \underbrace{I\left(W_1; \sum_{i=1}^K W_i \Bigg|Z_2,W_2,\{W_k,Z_k\}_{k \in \mathcal{T}}\right)}_{\overset{(\ref{ind_S})}{=}0}\notag\\
   &- \underbrace{I\left(W_1; X_1\Bigg|\sum_{i=1}^K W_i ,Z_2,W_2,\{W_k,Z_k\}_{k \in \mathcal{T}}\right)}_{\overset{(\ref{eq:security constraint})}{=}0}\label{thm3_eq5}\\
    \geq& L.
     \end{align}
The second term of (\ref{thm3_eq3}) equals zero since $\{X_k\}_{k\in [K]}$ is determined by $X_1,Z_2,W_2,\{W_k,X_k\}_{k \in \mathcal{U}},\{W_k,Z_k\}_{k \in \mathcal{T}}$. Given $\{X_k\}_{k\in [K]}$, one can obtain $\sum_{k=1}^K W_k$, and hence $W_1$ is known. In (\ref{thm3_eq5}), the first and second terms are justified by the independence of the inputs from the keys, whereas the third term equals zero due to the security requirement of User 2. Therefore, when $G=1$, the problem is infeasible.

When $G \geq K-T$, the problem becomes infeasible due to the security constraints. Recall that any user $k$ may collude with up to $T$ other users, which means that any group of $T+1$ users can freely share all the information they hold. Consider a groupwise key $S_{\mathcal{G}}$ associated with a subset $\mathcal{G} \subseteq [K]$ of size $|\mathcal{G}|=G$. By construction, this key is known to all users if there are up to $K-G+1$ users.
If $K-G+1 \le T+1$, then a coalition of $T+1$ users is sufficient to collectively access every user holding $S_{\mathcal{G}}$. In this case, the coalition can recover all groupwise keys in the system. Since the security requirement forbids any group of $T+1$ users from learning all keys, the problem cannot satisfy the security constraint.
Therefore, whenever $G \ge K-T$, any $T+1$ users can reconstruct all keys, and the problem is infeasible.

Also note that DSA is inherently infeasible when $K \le 2$ or $T > K-2$ as explained in~\cite{Zhang_Li_Wan_DSA}.
Therefore, we have implicitly assumed $K\ge  3$ and $T\le K-3$ in our problem.

\subsubsection{General Converse proof for $1<G\leq K-T-1$} \label{subsec:lemmas&corollaries}

This section presents several lemmas and corollaries that are central to the converse proof, based on which we derive tight bounds for the communication and groupwise rates.  
Following the approach used in the converse proof of Example 2, we prove each lemma sequentially.

The first lemma states that each message $X_k$ contains at least $L$ independent symbols, even when the inputs and keys of all other users are known.

\begin{lemma}
\label{lemma: H(Xk|(Wi,Zi) all other i)}
\emph{For any $u\in [K]$,  we have
\be 
\label{eq: H(Xk|(Wi,Zi) all other i), lemma}
H\left(X_u| \{W_k, Z_k\}_{k\in [K]\setminus \{u\}  } \right) \geq L.   
\ee 
}
\end{lemma}
\begin{IEEEproof}
For any other user $u\in[K]\setminus \{k\}$, $X_u$ must fully convey $W_u$ because it is the only source of that input, ensuring that other users can recover the sum. 
More formally, we have
\begin{align}
& H\left(X_u| \{W_k, Z_k\}_{k\in [K]\setminus \{u\}  } \right) \label{eq: proof H(Xk|(Wi,Zi) all other i), lemma}\\
\geq& I\left(X_u;W_u| \{W_k, Z_k\}_{k\in [K]\setminus \{u\}  } \right) \\
=&H\left(W_u| \{W_k, Z_k\}_{k\in [K]\setminus \{u\}  } \right)\notag\\
&-H\left(W_u| \{W_k, Z_k\}_{k\in [K]\setminus \{u\}  },X_u \right)\\
\overset{(\ref{message})}{=}&H\left(W_u\right)-H\left(W_u| \{W_k, Z_k,X_k\}_{k\in [K]\setminus \{u\}  },X_u \right)\label{lemma1pf1}\\
\overset{(\ref{eq:recovery constraint})}{=}&H\left(W_u\right)-\notag\\
&\underbrace{H\left(W_u\Bigg| \{W_k, Z_k\}_{k\in [K]\setminus \{u\}  },\{X_k\}_{k\in [K] },\sum_{k\in[K]}W_k \right)}_{=0}\label{lemma1pf2}\\
\overset{(\ref{eq:input independence})}{=}L,
\end{align}
where (\ref{lemma1pf1})  is because $X_k$ is a deterministic function of $W_k$ and $Z_k$ (see (\ref{message})). The second term in (\ref{lemma1pf2}) follows from the correctness constraint as $\sum_{k\in[K]}W_k$ can be recovered from $\{X_k\}_{k\in [K]\setminus\{u\}}$ and $W_{u},Z_{u}$. Moreover, the second term in (\ref{lemma1pf2}) vanishes because $W_u$ can be derived from $\{W_k\}_{k\in [K]\setminus\{u\}}$ and $\sum_{k\in[K]}W_k$.
\end{IEEEproof}

Let \(\mathcal T_{(k)} \subseteq [K]\setminus\{k\}\), \(|\mathcal T_{(k)}|\le K-3\), denote the set of users colluding with user \(k\), and define its complement (relative to \([K]\setminus\{k\}\)) by
$\overline{\mathcal T}_{(k)} \triangleq \big([K]\setminus\{k\}\big)\setminus \mathcal T_{(k)}.$
By Lemma~\ref{lemma: H(Xk|(Wi,Zi) all other i)}, the joint entropy of $\{X_i\}_{i\in \overline{\Tc}_{(k)}}$, conditioned on the inputs and keys of $\Tc_{(k)}$ and user $k$, is lower bounded by $|\overline{\Tc}_{(k)}|L$.

\begin{corollary}
\label{corollary1}
\emph{For any colluding user set $\Tc_{(k)} \subset [K]\bksl \{k\}$ and its complement $\overline{\Tc}_{(k)}$,  it holds that}
\be
\label{eq:corollary1}
H\left( \{X_i\}_{i\in \overline{\Tc}_{(k)} } \big| \{W_i,Z_i\}_{i\in\Tc_{(k)}}, W_k,Z_k       \right) \ge |\overline{\Tc}_{(k)} |L.
\ee 
\end{corollary}
\begin{IEEEproof}
Suppose $\overline{\Tc}_{(k)}=\big\{k_1, \cdots, k_{|\overline{\Tc}_{(k)}|}\big\}$. By the chain rule of entropy, we have
\begin{align}
& H\left( \{X_i\}_{i\in \overline{\Tc}_{(k)} } \big| \{W_i,Z_i\}_{i\in\Tc_{(k)}}, W_k,Z_k       \right )\notag\\
=&   \sum_{j=1}^{|\overline{\Tc}_{(k)}|} H\left( X_{k_j} \big| \{W_i,Z_i\}_{i\in\Tc_{(k)}\cup\{k\}  }, \{X_{k_i}\}_{i\in[1:j-1]}      \right)  \notag   \\
\ge & \sum_{j=1}^{|\overline{\Tc}_{(k)}|} H\Big( X_{k_j} \big| \{W_i,Z_i\}_{i\in\Tc_{(k)}\cup\{k\}\cup\{k_1,\cdots, k_{j-1}\}  },\notag\\
&   \{X_{k_i}\}_{i\in[1:j-1]}      \Big)\label{eq:step0,proof corollary1}\\  
 \overset{(\ref{message})}{=}& \sum_{j=1}^{|\overline{\Tc}_{(k)}|} H\left( X_{k_j} \big| \{W_i,Z_i\}_{i\in\Tc_{(k)}\cup\{k\}\cup\{k_1,\cdots, k_{j-1}\}  }\right)\label{eq:step1,proof corollary1}\\ 
 \ge &\sum_{j=1}^{|\overline{\Tc}_{(k)}|} H\left( X_{k_j} \big| \{W_i,Z_i\}_{i\in[K]\bksl \{k_j\}  }\right)\label{eq:step2,proof corollary1}\\ 
 \overset{(\ref{eq: H(Xk|(Wi,Zi) all other i), lemma})}{\ge}&  |\overline{\Tc}_{(k)}|L.
\end{align}
Here, (\ref{eq:step1,proof corollary1}) follows since $X_{k_i}$ is a deterministic function of $(W_{k_i},Z_{k_i})$ for all $i\in[1:j-1]$ (see (\ref{message})).  
Step (\ref{eq:step2,proof corollary1}) holds because $\Tc_{(k)}\cup\{k\}\cup\{k_1,\ldots,k_{j-1}\}\subseteq [K]\setminus\{k_j\}$ for each $j=1,\ldots,|\overline{\Tc}_{(k)}|$.  
Finally, Lemma~\ref{lemma: H(Xk|(Wi,Zi) all other i)} is invoked.
\end{IEEEproof}

The following lemma asserts that the message $X_k$ should remain independent of its associated input $W_k$ once we condition on the inputs and keys of all other users, namely ${W_u,Z_u, u\in [K]\setminus\{k\}}$. This property directly comes from the security requirement enforced at User $u$, ensuring that the information of $W_k$ is completely protected by the private key $Z_k$. Equivalently, we obtain that $I(X_k; W_k \mid W_{u}, Z_{u})=0$ for every $u\in [K]\setminus\{k\}$. The conditioning here indicates that User $u$ naturally possesses the pair $(W_u,Z_u)$, so these variables must be regarded as known when assessing the mutual information.
\begin{lemma}
\label{lemma: I(Xk;Wk|Wu,Zu)=0}
\emph{For any $k\in [K]$ and $u\in [K]\setminus \{k\}$ , we have
\begin{align} 
\label{eq:lemma: I(Xk;Wk|Wu,Zu)=0}
I\left(X_k; W_k |W_{u}, Z_{u}     \right)=0.
\end{align}
}
\end{lemma}
\begin{IEEEproof}
It holds that
\begin{align}
& I\left(X_k; W_k |W_{u}, Z_{u}     \right) 
 \le  I\left(X_k, \sum_{i=1}^KW_i; W_k |W_{u}, Z_{u}     \right)\notag\\
 =&   I\left(\sum_{i=1}^KW_i; W_k \Big|W_{u}, Z_{u}\right) + I\left(X_k; W_k \Big|W_{u}, Z_{u}, \sum_{i=1}^KW_i     \right)\\
\overset{(\ref{ind_S})}{\leq}&   I\left(\sum_{i=1}^KW_i; W_k \Big|W_{u}\right)\notag\\
&  + \underbrace{I\left( \{ X_i\}_{i\in [K]\bksl \{u\} };\{ W_k\}_{i\in [K]\bksl \{u\} } \Big|W_{u}, Z_{u}, \sum_{i=1}^KW_i     \right)}_{ \overset{(\ref{eq:security constraint})}{=}0 } \label{lemma2pf1}\\
 =&   H\left(\sum_{i=1}^KW_i\Big|W_{u}\right)- H\left(\sum_{i=1}^KW_i\Big| W_k ,W_{u}\right)\label{eq:step 0, proof I(Xk;Wk|Wu,Zu)=0}\\
 \overset{(\ref{eq:input independence})}{=}& H\left(\sum_{i\in[K]\bksl \{u\}  }W_i \right)- H\left(\sum_{i\in[K]\bksl \{u,k\}  }W_i \right)\label{eq:step 2, proof I(Xk;Wk|Wu,Zu)=0}\\
 \overset{(\ref{eq:input independence})}{=}& L-L=0,\label{eq:step 3, proof I(Xk;Wk|Wu,Zu)=0}
\end{align}
where in (\ref{lemma2pf1}), the first term follows from the independence between the inputs and the keys, while the second term is valid due to the security constraint when $\mathcal{T}_{(k)}=\emptyset$. Furthermore, (\ref{eq:step 2, proof I(Xk;Wk|Wu,Zu)=0}) holds because of the input independence property (see (\ref{eq:input independence})) together with the assumption $K\geq 3$.
\end{IEEEproof}

\begin{lemma}
\label{lemma: I((Xi)_Tkc;(Wi)_Tkc|Wk,Zk,(Wi,Zi)_Tk)=L}
\emph{
For any $k\in [K]$, any colluding user set $\Tc_{(k)} \subset [K]\bksl \{k\}$ and its complement $\overline{\Tc}_{(k)}$, the following equality holds:}
\begin{align}
\label{eq:I((Xi)_Tkc;(Wi)_Tkc|Wk,Zk,(Wi,Zi)_Tk)=L, lemma}
 & I\left(\{X_i\}_{i\in \overline{\Tc}_{(k)}}; 
\{W_i\}_{i\in \overline{\Tc}_{(k)}} \big|\{W_i,Z_i\}_{i\in \Tc_{(k)}}, W_k,Z_k\right)\leq L.
\end{align} 
\end{lemma}
\begin{IEEEproof}
We have
\begin{align}
& I\left(\{X_i\}_{i\in \overline{\Tc}_{(k)}}; 
\{W_i\}_{i\in \overline{\Tc}_{(k)}} \big|\{W_i,Z_i\}_{i\in \Tc_{(k)}   }, W_k,Z_k\right) \notag\\
\leq& I\left(\{X_i\}_{i\in \overline{\Tc}_{(k)}}, \sum_{i\in \overline{\Tc}_{(k)}   } W_i; 
\{W_i\}_{i\in \overline{\Tc}_{(k)}} \bigg|\right.\notag\\
&\left.\{W_i,Z_i\}_{i\in \Tc_{(k)} \cup  \{k\} }\right)\\
 =& I\left( 
\sum_{i\in \overline{\Tc}_{(k)}   } W_i ;\{W_i\}_{i\in \overline{\Tc}_{(k)}} \bigg|\{W_i,Z_i\}_{i\in \Tc_{(k)} \cup  \{k\} }\right)\notag\\
& + I\left(\{X_i\}_{i\in \overline{\Tc}_{(k)}}; 
\{W_i\}_{i\in \overline{\Tc}_{(k)}} \bigg|\right.\notag\\
&\left.\{W_i,Z_i\}_{i\in \Tc_{(k)}\cup  \{k\} }, \sum_{i\in \overline{\Tc}_{(k)}   } W_i\right)\label{eq:step 0, proof I((Xi)_Tkc;(Wi)_Tkc|Wk,Zk,(Wi,Zi)_Tk)=L, lemma}\\
 \leq& H\left( \sum_{i\in \overline{\Tc}_{(k)}   } W_i \bigg|\{W_i,Z_i\}_{i\in \Tc_{(k)} \cup  \{k\} }\right) \notag\\
&  - \underbrace{H\left( \sum_{i\in \overline{\Tc}_{(k)}   } W_i \bigg|\{W_i,Z_i\}_{i\in \Tc_{(k)}\cup  \{k\} }, \{W_i\}_{i\in \overline{\Tc}_{(k)}}\right)}_{=0}\notag\\
& + \underbrace{I\left(\{X_i\}_{i\in [K]\setminus \{k\}}; 
\{W_i\}_{i\in [K]\setminus \{k\}} \bigg|\right.}_{\overset{(\ref{eq:security constraint})}{=}0}\notag\\
&\underbrace{\left.\{W_i,Z_i\}_{i\in \Tc_{(k)}\cup  \{k\} }, \sum_{i\in \overline{\Tc}_{(k)}   } W_i\right)}_{\overset{(\ref{eq:security constraint})}{=}0}\label{lemma3pf1}\\
\overset{(\ref{ind_S})}{=}&H\left( \sum_{i\in \overline{\Tc}_{(k)}   } W_i \right) \label{lemma3pf2}\\
\overset{(\ref{eq:input independence})}{=}&L,
\end{align}
where in (\ref{lemma3pf1}), the second term becomes zero since $\sum_{i\in \overline{\Tc}{(k)}} W_i$ is fully determined by $\{W_i\}_{i\in \overline{\Tc}_{(k)}}$. The third term vanishes because of the security constraint (see (\ref{eq:security constraint})). Finally, (\ref{lemma3pf2}) follows from the independence between the inputs and the keys.
\end{IEEEproof}

Lemma~\ref{lemma: I((Xi)_Tkc;(Wi)_Tkc|Wk,Zk,(Wi,Zi)_Tk)=L} indicates that, from the viewpoint of any user $k$, even if it colludes with users in $\Tc{(k)}\subset [K]\setminus\{k\}$, the only information that can be obtained about the inputs $\{W_i\}_{i\in \overline{\Tc}{(k)}}$ is their sum. This captures the essential security requirement for user $k$: it should not gain any knowledge beyond the sum of these inputs and what is already accessible through the colluding users, no matter which subset $\Tc_{(k)}$ is involved.

The following lemma states that for any user $k \in [K]$, the joint entropy of the individual keys in $\overline{\Tc}{(k)}$, conditioned on all keys in $\Tc{(k)} \cup \{k\}$, is no less than $(K-T-2)L$.
\begin{lemma}
\label{lemma:H((Zi)_Tkc|(Zi,Wi)_Tk, Wk,Zk)>=(K-|Tkc|-2)L}
\emph{
For any set of colluding users $\Tc_{(k)}\in [K]\bksl\{k\}$ and its complement $\overline{\Tc}_{(k)}=( [K]\bksl\{k\})\bksl \Tc_{(k)}$, we have}
\be
\label{eq:H((Zi)_Tkc|(Zi,Wi)_Tk, Wk,Zk)>=(K-|Tkc|-2)L, lemma}
H\left( \{Z_i\}_{ i\in\overline{\Tc}_{(k)}  } \big| \{Z_i\}_{i\in \Tc_{(k)}},Z_k   \right)
\geq \left(K-T-2\right)L. 
\ee
\end{lemma}

\begin{IEEEproof}
It holds that
\begin{align}
& H\left( \{Z_i\}_{ i\in\overline{\Tc}_{(k)}  } \big| \{Z_i\}_{i\in \Tc_{(k)}},Z_k   \right)\notag\\
 \overset{(\ref{ind_S})}{=}&
H\left( \{Z_i\}_{ i\in\overline{\Tc}_{(k)}  } \big|\{W_i\}_{ i\in\overline{\Tc}_{(k)}  } , \{Z_i,W_i\}_{i\in \Tc_{(k)}},W_k,Z_k 
 \right)\label{eq:step0,proof,H((Zi)_Tkc|(Zi,Wi)_Tk, Wk,Zk)>=(K-|Tkc|-2)L, lemma}\\
 \geq & I\Big( \{Z_i\}_{ i\in\overline{\Tc}_{(k)}  }; \{X_i\}_{ i\in\overline{\Tc}_{(k)}  }  \big|\{W_i\}_{ i\in\overline{\Tc}_{(k)}  } , \{Z_i,W_i\}_{i\in \Tc_{(k)}},\notag\\
&\quad W_k,Z_k \Big)\\
 =& H\left( \{X_i\}_{ i\in\overline{\Tc}_{(k)}  }  \big|\{W_i\}_{ i\in\overline{\Tc}_{(k)}  } , \{Z_i,W_i\}_{i\in \Tc_{(k)}}, W_k,Z_k \right)-\notag\\
& \underbrace{H\left( \{X_i\}_{ i\in\overline{\Tc}_{(k)}  }  \big|\{W_i,Z_i\}_{ i\in\overline{\Tc}_{(k)}  } , \{Z_i,W_i\}_{i\in \Tc_{(k)}}, W_k,Z_k \right)}_{\overset{(\ref{message})}{=}0  }\label{lemma4pf1}\\
 = &
H\left( \{X_i\}_{ i\in\overline{\Tc}_{(k)}  }  \big| \{W_i,Z_i\}_{i\in \Tc_{(k)}}, W_k,Z_k \right)  \notag\\
&- I\left( \{X_i\}_{ i\in\overline{\Tc}_{(k)}  }; \{W_i\}_{ i\in\overline{\Tc}_{(k)}  }  \big| \{W_i,Z_i\}_{i\in \Tc_{(k)}}, W_k,Z_k \right) \\
 &\overset{(\ref{eq:corollary1}),(\ref{eq:I((Xi)_Tkc;(Wi)_Tkc|Wk,Zk,(Wi,Zi)_Tk)=L, lemma})}{\ge }
|\overline{\Tc}_{(k)}|L-L\label{eq:step1,proof,H((Zi)_Tkc|(Zi,Wi)_Tk, Wk,Zk)>=(K-|Tkc|-2)L, lemma}\\
 =& \left(K-|\Tc_{(k)}|-2\right)L\label{lemma4pf2}\\
\geq &\left(K-T-2\right)L,
\end{align}
where (\ref{eq:step0,proof,H((Zi)_Tkc|(Zi,Wi)_Tk, Wk,Zk)>=(K-|Tkc|-2)L, lemma}) follows from the independence between the inputs and the keys. In (\ref{lemma4pf1}), the second term is zero because $\{X_i\}_{i\in\overline{\Tc}{(k)}}$ is completely determined by $\{W_i\}_{i\in\overline{\Tc}{(k)}}$. In (\ref{eq:step1,proof,H((Zi)_Tkc|(Zi,Wi)_Tk, Wk,Zk)>=(K-|Tkc|-2)L, lemma}), we apply Corollary~\ref{corollary1} and Lemma~\ref{lemma: I((Xi)_Tkc;(Wi)_Tkc|Wk,Zk,(Wi,Zi)_Tk)=L}. (\ref{lemma4pf2}) holds because $|\overline{\Tc}{(k)}| + |\Tc{(k)}| = K-1$. The last step follows from the fact that $\Tc_{(k)}$ can be any subset with $|\Tc_{(k)}| \le T$. By choosing $|\Tc_{(k)}| = T$, we have $(K - |\Tc_{(k)}| - 2)L = (K - T - 2)L$. Therefore, the proof of Lemma~\ref{lemma:H((Zi)_Tkc|(Zi,Wi)_Tk, Wk,Zk)>=(K-|Tkc|-2)L} is complete. 
\end{IEEEproof}

With the lemmas established above, we are now prepared to derive the lower bounds on the communication and key rates. Because these bounds coincide with the achievable rates presented in Section~\ref{sec: ach scheme}, the optimality of the proposed scheme is thereby confirmed.

\subsubsection{Proof of \Comm Rate $\rx \ge 1$}
\label{subsubsec:proof of Rx>=1}
Consider any user $k\in[K]$. 
A straightforward application of Lemma~\ref{lemma: H(Xk|(Wi,Zi) all other i)} gives the following bound on the \comm rate:
\begin{align}
L_X \ge H(X_k) & \ge H\left(X_k |\{W_i, Z_i\}_{i\in [K]\bksl \{k\}   } \right)\overset{(\ref{eq: H(Xk|(Wi,Zi) all other i), lemma})}{\ge } L\\
  \Rightarrow \rx  & \eqdef {L_X}/{L} \geq  1.
\end{align}

\subsubsection{Proof of Groupwise Key Rate $R_S \ge \cfrac{K-T-1}{\binom{K-T}{G}}$}
\label{subsubsec:proof of Rz>=1}
From lemma~\ref{lemma:H((Zi)_Tkc|(Zi,Wi)_Tk, Wk,Zk)>=(K-|Tkc|-2)L}, we have
\begin{align}
&\left(K-2-|\Tc_{(k)}|\right)L\notag\\
\overset{(\ref{eq:H((Zi)_Tkc|(Zi,Wi)_Tk, Wk,Zk)>=(K-|Tkc|-2)L, lemma})}{\leq}& H\left( \{Z_i\}_{ i\in\overline{\Tc}_{(k)}  } \big| \{Z_i\}_{i\in \Tc_{(k)}},Z_k   \right)\label{rspf1}\\
    \overset{(\ref{Z_individual})}{=}& H\left((S_\mathcal{G})_{\mathcal{G} \in \binom{[K]}{G},\mathcal{G} \cap \overline{\mathcal{T}}_{(k)} \neq \emptyset} \Big| (S_\mathcal{G})_{\mathcal{G} \in \binom{[K]}{G},\mathcal{G} \cap (\mathcal{T}_{(k)} \cup \{k\})\neq \emptyset}\right) \label{rspf2}\\
    \overset{(\ref{ind_S})}{=}& H\left((S_\mathcal{G})_{\mathcal{G} \in \binom{[K]}{G},\mathcal{G} \cap \overline{\mathcal{T}}_{(k)} \neq \emptyset,\mathcal{G} \cap (\mathcal{T}_{(k)} \cup \{k\}) = \emptyset}\right) \label{rspf3}\\
    =& H\left((S_\mathcal{G})_{\mathcal{G} \in \binom{\overline{\mathcal{T}}_{(k)}}{G}}\right)\\
    \overset{(\ref{ind_S})}{=}& \binom{|\overline{\mathcal{T}}_{(k)}|}{G} \times L_S,\label{rspf4}\\
\Rightarrow & R_S= \frac{L_S}{L} \geq \cfrac{K-T-2}{\binom{K-T-1}{G}}. \label{eq:e5}
\end{align}
In (\ref{rspf1}), we apply Lemma~\ref{lemma:H((Zi)_Tkc|(Zi,Wi)_Tk, Wk,Zk)>=(K-|Tkc|-2)L}. In (\ref{rspf2}), each key variable $Z_k$ is replaced by the corresponding groupwise keys (see (\ref{Z_individual})). Equations (\ref{rspf3}) and (\ref{rspf4}) follow from the independence of the groupwise keys (see (\ref{ind_S})).

\section{Conclusion}
\label{sec:conclusion & future directions}

In this paper, we studied the decentralized secure aggregation (DSA) problem in a fully-connected \decen network under groupwise key structures and user collusion. We established an exact capacity region characterizing the minimum communication overhead and secret key size required to securely compute the sum of user inputs, showing that DSA is infeasible when the group size is either too small or too large, and deriving tight achievable and converse bounds for all feasible regimes. The proposed key-neutralization based construction attains the optimal rate region and reveals how groupwise key structures fundamentally govern both feasibility as well as \comm and secret key generation efficiency. Our results extend secure aggregation beyond centralized architectures, and provide practical guidance for designing secure and communication-efficient federated learning systems in decentralized topologies.

\bibliographystyle{IEEEtran}
\bibliography{references_secagg.bib}

\end{document}